\documentclass[a4paper,11pt]{article}
\usepackage[dvipdfmx]{graphicx}
\usepackage{amsfonts}
\usepackage{physics,amsmath}
\usepackage{amssymb}
\usepackage{mathtools}
\usepackage{colortbl}
\usepackage[table]{xcolor}
\usepackage{tikz}
\usepackage[compat=1.1.0]{tikz-feynman}
\usepackage{cite}
\usepackage{here}
\usepackage{hyperref}
\usepackage{mathrsfs}
\usepackage{algpseudocode}
\usepackage{ulem}
\usepackage{bm}
\usepackage{comment}
\numberwithin{equation}{section}
\usepackage[left=2cm,right=2cm,top=3.5cm,bottom=3.5cm]{geometry}

\allowdisplaybreaks[1]
\hypersetup{
	colorlinks=true,
	linkcolor=ceruleanblue,
	filecolor=ceruleanblue,      
	urlcolor=ceruleanblue,
	citecolor=ceruleanblue,
}
\definecolor{ceruleanblue}{rgb}{0.0, 0.2, 0.6}
\newcommand{\de}{{\rm d}}
\usepackage[table]{xcolor}
\usepackage{tikz}
\usepackage[compat=1.1.0]{tikz-feynman}
\tikzset{graviton/.style={decorate, decoration={snake, amplitude=.4mm, segment length=1.5mm, pre length=.5mm, post length=.5mm}, double}}

\newcommand{\ExternalSource}{
\begin{tikzpicture}[baseline]
\begin{feynman}
    \vertex[dot] (a0);
    \vertex[below=1cm of a0] (p1);
    \vertex[above=1cm of a0] (p2);
    \vertex[right=0.4cm of a0, blob] (gblob){};			
    \vertex [crossed dot, minimum size=0.25cm, label=0:$j_{\Hat{\ell}}(\omega r)$, right=1.5cm of p1] (b1) {};
    \vertex[right=1.5cm of p2] (b2);
    \vertex[above=0.7cm of a0] (g1);    \vertex[above=0.4cm of a0] (g2);
    \vertex[right=0.05cm of a0] (gdtos){$\vdots$};
    \vertex[below=0.7cm of a0] (gN);
    \diagram*{
	(p1) -- [double,double distance=0.5ex] (p2),
	(g1) -- [photon] (gblob),
	(g2) -- [photon] (gblob),
	(gN) -- [photon] (gblob),
	(b1) -- (gblob) -- [photon](b2),
        };
\end{feynman}
\end{tikzpicture}}

\newcommand{\InducedResponse}{
\begin{tikzpicture}[baseline]
\begin{feynman}
    \vertex[dot] (a0);
    \vertex [square dot, minimum size=0.25cm, label=180:$\overline{\cal F}_{\hat{\ell}}^{\rm B}(\omega)$, below=0.1cm of a0] (a) {};
    \vertex[below=1.0cm of a0] (p1);
    \vertex[above=1.0cm of a0] (p2);
    \vertex [crossed dot, minimum size=0.25cm, label=0:$j_{\Hat{\ell}}(\omega r)$, right=1cm of p1] (b1) {};
    \vertex[right=1.0cm of p2] (b2);
    \diagram*{
	(p1) -- [double,double distance=0.5ex] (a)  -- [double,double distance=0.5ex] (p2),,
	(b2) --[photon] (a)  --  (b1),
    };
\end{feynman}
\end{tikzpicture}}

\newcommand{\InducedResponseCorrected}{
\begin{tikzpicture}[baseline]
\begin{feynman}
    \vertex[dot] (a0);
    \vertex [square dot, minimum size=0.25cm, label=180:$\overline{\cal F}_{\hat{\ell}}^{\rm B}(\omega)$, below=0.3cm of a0] (a) {};
    \vertex[below=1.0cm of a0] (p1);
    \vertex[above=1.1cm of a0] (p2);
    \vertex [crossed dot, minimum size=0.25cm, label=0:$j_{\Hat{\ell}}(\omega r)$, right=1.4cm of p1] (b1) {};
    \vertex[right=1.5cm of p2] (b2);
    \vertex[below=0.1cm of a0] (g1);
    \vertex[above=0.8cm of g1] (g2);
    \vertex[below=0.2cm of g2] (gdotsaux);
    \vertex[right=0.00cm of gdotsaux] (gdtos){$\vdots$};
    \vertex[above=0.4cm of gdotsaux] (gN);
    \vertex[above=0.5cm of a0] (gblobaux);
    \vertex[right=0.5cm of gblobaux, blob] (gblob){};
    \diagram*{
	(p1) -- [double,double distance=0.5ex] (a)  -- [double,double distance=0.5ex] (p2),
        (b2) --[photon] (gblob) --[photon] (a)  -- (b1),
	(g1) -- [photon] (gblob),
	(g2) -- [photon] (gblob),
	(gN) -- [photon] (gblob),
    };
\end{feynman}
\end{tikzpicture}}

\let\originalleft\left
\let\originalright\right
\renewcommand{\left}{\mathopen{}\mathclose\bgroup\originalleft}
\renewcommand{\right}{\aftergroup\egroup\originalright}

\date{\today}

\begin{document}
\begin{flushright} {\footnotesize YITP-25-176, RESCEU-25/25, IPMU25-0052, RIKEN-iTHEMS-Report-25}  \end{flushright}

\begin{center}
\LARGE{\bf Dynamical Tidal Response of Non-rotating Black Holes:\\
Connecting the MST Formalism and Worldline EFT}
\\[1cm] 

\large{
Hajime Kobayashi$^{\,\rm a}$, Shinji Mukohyama$^{\,\rm a, \rm b, \rm c}$, Naritaka Oshita$^{\,\rm a, \rm d, \rm e}$, Kazufumi Takahashi$^{\,\rm f, \rm a}$,\\
and Vicharit Yingcharoenrat$^{\,\rm g, \rm c}$}
\\[0.5cm]

\small{\textit{$^{\rm a}$
Center for Gravitational Physics and Quantum Information, Yukawa Institute for Theoretical Physics, 
\\ Kyoto University, 606-8502, Kyoto, Japan}}
\vspace{.2cm}

\small{\textit{$^{\rm b}$
Research Center for the Early Universe (RESCEU), Graduate School of Science, The University of Tokyo, Hongo 7-3-1, Bunkyo-ku, Tokyo 113-0033, Japan}}
\vspace{.2cm}

\small{
\textit{$^{\rm c}$
Kavli Institute for the Physics and Mathematics of the Universe (WPI), The University of Tokyo Institutes for Advanced Study (UTIAS), The University of Tokyo, Kashiwa, Chiba 277-8583, Japan}}
\vspace{.2cm}

\small{
\textit{$^{\rm d}$
The Hakubi Center for Advanced Research, Kyoto University, Yoshida Ushinomiyacho, Sakyo-ku, Kyoto 606-8501, Japan}}
\vspace{.2cm}

\small{
\textit{$^{\rm e}$
RIKEN iTHEMS, Wako, Saitama, 351-0198, Japan}}
\vspace{.2cm}

\small{
\textit{$^{\rm f}$
Department of Physics, College of Humanities and Sciences, Nihon University, Tokyo 156-8550, Japan}}
\vspace{.2cm}

\small{
\textit{$^{\rm g}$
High Energy Physics Research Unit, Department of Physics, Faculty of Science, Chulalongkorn University, Pathumwan, Bangkok 10330, Thailand}}
\vspace{.2cm}
\end{center}

\vspace{0.3cm} 

\begin{abstract}\normalsize
The response of a black hole (BH) to tidal forces encodes key information about the underlying gravitational theory and affects the waveform of gravitational waves emitted during binary inspiral processes.
In this paper, we analyze the dynamical tidal response of static and spherically symmetric BHs in a low-frequency regime within general relativity (GR), based on a matching between the Mano-Suzuki-Takasugi (MST) methods for an analytical approach to BH perturbations and the worldline effective field theory (EFT) for an efficient and unified computation of the binary dynamics within the post-Newtonian regime.
We show that the renormalized tidal response function is subject to inevitable ambiguities associated with the choice of renormalization scheme and with the initial condition of the renormalization flow equation.
Once these ambiguities are fixed, we obtain scheme-dependent dynamical tidal Love numbers.
We also discuss possible extensions of our formalism, including generic non-rotating compact objects (e.g., neutron stars) in GR and BHs in theories beyond GR.
\end{abstract}

\vspace{0.3cm} 

\vspace{2cm}

\newpage
{
\hypersetup{linkcolor=black}
\tableofcontents
}

\flushbottom
\clearpage
\section{Introduction}\label{sec:Introduction}
Gravitational waves (GWs) from a coalescence of binary compact objects provide a powerful and insightful probe to study fundamental properties of gravity and compact objects in the strong-field regime.
In particular, GWs emitted from a late inspiral are one of the most promising observational targets.
As the number of merger events detected by the LIGO-Virgo-KAGRA collaboration~\cite{LIGOScientific:2016aoc,LIGOScientific:2018mvr,LIGOScientific:2020ibl,LIGOScientific:2021usb,KAGRA:2021vkt} continues to grow, and as next generation detectors---such as the Einstein Telescope~\cite{Punturo:2010zz,ET:2019dnz,Branchesi:2023mws,Abac:2025saz}, Cosmic Explorer~\cite{Reitze:2019iox,Evans:2021gyd}, and Laser Interferometer Space Antenna~\cite{LISA:2017pwj,LISA:2024hlh}---promise to deliver GW observations with higher signal-to-noise ratios, we are entering an era of precision tests of strong gravity.
This requires an accurate and precise waveform modeling, which includes a precise understanding of conservative and dissipative dynamics of a two-body system originating from tidal effects (see, e.g., Refs.~\cite{Flanagan:2007ix,Blanchet:2013haa,Henry:2020ski,Kalin:2020lmz,Buonanno:2022pgc,Goldberger:2022ebt,Chia:2024bwc,Shterenberg:2024tmo}).

During the initial inspiral phase, where the orbital separation of the binary is much larger than the size of the binary components, they are well approximated by point particles~\cite{Blanchet:2013haa}.
However, as the objects spiral toward each other, tidal effects become significant, producing a perturbative deformation and influencing the orbital evolution through both conservative and dissipative mechanisms.
The internal properties of the objects and/or the underlying gravitational theory are reflected in {\it tidal response functions}.
Since the wavelengths of tidal perturbations during the inspiral phase are much longer than the scale of the internal dynamics, the coefficients in the low-frequency expansion of the tidal response functions can be constrained by GW observations.

In the post-Newtonian (PN) regime of the inspiral system, the leading tidal effects conserving the orbital energy enter at 5PN order~\cite{Flanagan:2007ix,Damour:2012yf,Favata:2013rwa}.
Using the data from the GW event~GW170817, the 5PN term in the GW phase was constrained for the first time~\cite{LIGOScientific:2017vwq,LIGOScientific:2018cki,LIGOScientific:2018hze}.
The leading conservative tidal responses are characterized by the so-called {\it (static) tidal Love numbers} (TLNs)~\cite{Hinderer:2007mb,Damour:2009vw,Binnington:2009bb,Kol:2009mj,Kol:2011vg}, which quantify the deformability of an object in a static tidal field.
In addition to the conservative tidal effects, there are dissipative tidal responses that cause tidal heating and affect the gravitational waveform at 4PN order for non-rotating objects and at 2.5PN order for spinning objects~\cite{Poisson:1994yf,Tagoshi:1997jy, Alvi:2001mx,Poisson:2004cw}.
Such dissipative effects are quantified by the so-called {\it tidal dissipation numbers} (TDNs)~\cite{Ivanov:2022hlo,Saketh:2022xjb,Katagiri:2023yzm,Katagiri:2024fpn,Chia:2024bwc}.
These coefficients in the tidal response function encode the physical properties of the objects, which can provide valuable insight into the internal states and dynamics of neutron stars within general relativity (GR)~\cite{Flanagan:2007ix,Hinderer:2007mb,Damour:2009vw,Binnington:2009bb,Damour:2012yf,Ghosh:2023vrx,Ripley:2023qxo,Saketh:2024juq,Katagiri:2025qze,Silvestrini:2025lbe}, gravity beyond GR through black holes (BHs)~\cite{Cardoso:2017cfl,Cardoso:2018ptl,Chakravarti:2018vlt,Cardoso:2019vof,DeLuca:2022tkm,DeLuca:2022xlz,Katagiri:2023umb,Hirano:2024pmk,Barura:2024uog,Katagiri:2024fpn,Barbosa:2025uau,Cano:2025zyk,Coviello:2025pla,Kobayashi:2025swn,Bhattacharyya:2025slf} or neutron stars~\cite{Pani:2014jra,Yazadjiev:2018xxk,Saffer:2021gak,Brown:2022kbw,Creci:2023cfx,Bernard:2023eul,Lam:2024azd,Creci:2024wfu,Diedrichs:2025vhv,Kobayashi:2025bdh}, environmental fields surrounding BHs~\cite{Cardoso:2019upw,DeLuca:2021ite,Cardoso:2021wlq,DeLuca:2022xlz,Katagiri:2023yzm,Brito:2023pyl,Capuano:2024qhv,DeLuca:2024uju,Cannizzaro:2024fpz,DeLuca:2025bph}, and the possible detection of hypothetical exotic compact objects~\cite{Uchikata:2016qku,Maselli:2017cmm,Cardoso:2017cfl,Addazi:2018uhd,Maselli:2018fay,Cardoso:2019rvt,Cardoso:2019nis,Datta:2019epe,Maggio:2021uge,Chakraborty:2023zed,Silvestrini:2025lbe}.\footnote{See Ref.~\cite{Silvestrini:2025lbe} for a unified formalism of TLNs and TDNs for generic compact objects based on the membrane paradigm.}

Although these coefficients are sufficient to describe the tidal response at the leading PN order, it is necessary to take into account subleading terms for precise waveform modeling.
These subleading terms originate from finite-frequency corrections or non-linear effects.
As for the finite-frequency corrections, they become significant in the late stage of the inspiral, when the relative velocity of the two bodies increases.
In the case of a non-rotating body, conservative dynamical tidal effects, quantified by the so-called {\it dynamical tidal Love numbers} (dTLNs), first contribute at 8PN order~\cite{Hinderer:2007mb,Hinderer:2016eia,Suvorov:2024cff}, while dissipative effects enter already at 7PN order.\footnote{For a neutron star, dTLNs quantify the oscillatory tidal deformation induced by an external perturbation within a low-frequency expansion and are widely used in waveform modeling of neutron-star binary mergers~\cite{Chakrabarti:2013lua,Hinderer:2016eia,Steinhoff:2016rfi,Steinhoff:2021dsn,Suvorov:2024cff,HegadeKR:2024agt}. See also Ref.~\cite{Miao:2025utd} for an alternative formulation describing the relativistic excitation of internal modes.}
Also, regarding the corrections due to the non-linearities of the Einstein equation, the effects of quadratic TLNs enter at 8PN order~\cite{Pitre:2025qdf}.

As for BHs in GR, it was shown that the static TLNs vanish for all modes with generic multipoles~$\ell\ge 2$~\cite{Damour:2009vw,Binnington:2009bb,Kol:2011vg,Gurlebeck:2015xpa,Poisson:2014gka,LeTiec:2020spy,Hui:2020xxx,Chia:2020yla,LeTiec:2020bos,Charalambous:2021mea,Bhatt:2023zsy}, which can be understood in terms of hidden ladder symmetries~\cite{Charalambous:2021kcz,Hui:2021vcv,BenAchour:2022uqo,Hui:2022vbh,Charalambous:2022rre,Katagiri:2022vyz,Berens:2022ebl,Sharma:2024hlz}.
From this fact, it is interesting to study whether dTLNs and quadratic TLNs vanish or not at both theoretical and observational levels.
Non-linear corrections to TLNs of BHs were previously studied in Refs.~\cite{Poisson:2009qj,Gurlebeck:2015xpa,Poisson:2020vap,Poisson:2021yau,DeLuca:2023mio,Riva:2023rcm,Iteanu:2024dvx,Combaluzier-Szteinsznaider:2024sgb,Kehagias:2024rtz,Gounis:2024hcm,Pitre:2025qdf,Lupsasca:2025pnt}, suggesting that the non-linear TLNs vanish for Schwarzschild BHs in GR.
In particular, it was shown in Refs.~\cite{Combaluzier-Szteinsznaider:2024sgb,Kehagias:2024rtz,Gounis:2024hcm,Lupsasca:2025pnt} that the non-linear TLNs vanish at all orders of axisymmetric metric perturbations from the viewpoint of the hidden symmetry.
On the other hand, dTLNs were calculated within GR for non-rotating BHs and neutron stars in Refs.~\cite{Chakrabarti:2013lua,Poisson:2020vap,Saketh:2023bul,Katagiri:2024wbg,HegadeKR:2024agt,Pitre:2025qdf,Chakraborty:2025wvs,HegadeKR:2025qwj}, but there are subtle unresolved issues regarding the definition of dTLNs with a logarithmic running.\footnote{In the case of Kerr BHs, the real part of the response function contains a term linear in the frequency with a logarithmic running~\cite{Charalambous:2021mea,Perry:2023wmm,Perry:2024vwz,Bhatt:2024yyz,Bhatt:2024rpx}, which is also referred to as dTLNs~\cite{Perry:2023wmm,Perry:2024vwz,Bhatt:2024yyz,Bhatt:2024rpx}. In the Schwarzschild limit, this linear-in-frequency contribution vanishes, and the logarithmic running instead appears at quadratic order in the frequency~\cite{Perry:2023wmm}.}

In this work, we revisit these problems through matching between the Mano-Suzuki-Takasugi (MST) solution~\cite{Mano:1996gn,Mano:1996mf,Mano:1996vt} (see also Ref.~\cite{Sasaki:2003xr} for a review) of Regge-Wheeler equation~\cite{Regge:1957td} and the worldline effective field theory (EFT)~\cite{Goldberger:2004jt,Goldberger:2005cd,Goldberger:2006bd,Porto:2016pyg,Levi:2018nxp,Goldberger:2022ebt,Goldberger:2022rqf}.
We argue that the renormalized tidal response function is subject to inevitable ambiguities associated with the choice of renormalization scheme and with the initial condition of the renormalization flow equation.
Once these ambiguities are fixed, we obtain scheme-dependent dTLNs.

The rest of the paper is organized as follows.
In Sec.~\ref{sec:Review-setup}, we provide a brief review of the formulation of tidal response.
In Sec.~\ref{sec:Matching}, we discuss the matching between the worldline EFT and the MST solution, as well as the ambiguities associated with the renormalization procedure.
In Sec.~\ref{sec:Extensions}, we discuss possible extensions of our formalism, including generic non-rotating compact objects (e.g., neutron stars) in GR and BHs in theories beyond GR.
Finally, in Sec.~\ref{sec:Conclusions}, we draw our conclusions.

\paragraph{Notations}
Throughout the paper, we use the unit~$\hbar=c=G=1$, where $G$ denotes the Newton's constant.
We use the mostly-plus signature of the metric.
The spacetime dimension is denoted by $d$; we set $d=4-\epsilon$ when applying dimensional regularization, and $d=4$ otherwise.
We use $\mu,\nu,\rho,\cdots$ for $d$-dimensional spacetime indices, $i,j,k,\cdots$ for $D=d-1$ spatial indices, and $a,b,c,\cdots$ denote angular indices on the $(d-2)$-dimensional sphere.
Our convention for the Fourier transform with respect to a time coordinate (e.g., $t$) is $\Psi(t,\bm{x})=\int\frac{\de\omega}{2\pi}e^{-i\omega t}\tilde{\Psi}(\bm{x};\omega)$.
We use the notation~$x^L\equiv x^{i_1}x^{i_2}\cdots x^{i_\ell}$ to denote the product of $\ell$ spatial coordinates, where $L=i_1 i_2\cdots i_\ell$ is a multi-index.
The symbols~$(\cdots)$ and $[\cdots]$ denote the symmetrization and anti-symmetrization over the enclosed indices, i.e., $A_{(\mu}B_{\nu)}\equiv \frac{1}{2}(A_\mu B_\nu+A_\nu B_\mu)$ and $A_{[\mu}B_{\nu ]}\equiv \frac{1}{2}(A_\mu B_\nu-A_\nu B_\mu)$, respectively.
In addition, $n_{\langle L\rangle}$ denotes the symmetric trace-free (STF) part of a multi-index quantity~$n_{L}$.

\section{Review of setup}\label{sec:Review-setup}
In this section, we present the setup for the relativistic formulation of dynamical tidal response based on the matched asymptotic expansion approach~(see, e.g., Refs.~\cite{Detweiler:2005kq,Taylor:2008xy,Poisson:2018qqd,Poisson:2020vap,HegadeKR:2024agt,Katagiri:2024wbg}).
First, in Sec.~\ref{ssec:scale}, we introduce the separation of scales during the inspiral phase, which defines the so-called body zone and PN zone.
We then present the analytical descriptions of these two zones in Secs.~\ref{ssec:MST} and \ref{ssec:worldline-EFT}, respectively.

\subsection{Separation of scales during inspiral process}\label{ssec:scale}
The description of an inspiralling system relies on the matched asymptotic expansion between the neighborhood of the compact objects and the PN metric describing the external region.
Suppose we focus on one of the binary constituents, whose Schwarzschild radius is $r_g$.
We assume that (i)~the inspiralling system has a large separation, $r_{\rm orb} \gg r_g$, with $r_{\rm orb}$ the orbital separation, and (ii)~the tidal environment evolves adiabatically, such that $v_{\rm orb} \equiv \omega r_{\rm orb} \ll 1$, with $\omega$ the typical orbital angular frequency and $v_{\rm orb}$ the relative velocity of the binary during the inspiral.
Then, the {\it body zone} is defined by $r_g\lesssim r\ll r_{\rm orb}$, where gravity is strong and described by full relativistic theories of perturbations.
Also, the {\it PN zone} is defined by $r_g\ll r \ll \omega^{-1}$, where gravity is weak and described by the PN metric.
The overlap region~$r_g\ll r\ll r_{\rm orb}\,(\ll \omega^{-1})$ is called the {\it buffer zone}, where the body zone metric and PN metric are matched.
This matching allows us to derive the overall spacetime dynamics, which in turn enables us to formulate tidal responses that link the theory of compact objects with the inspiral dynamics observed in GWs.\footnote{We do not consider the more distant region~$r\gtrsim\omega^{-1}$, which is called the wave (or radiative) zone, since it is irrelevant for the definition of the tidal response itself. Of course, this region must also be taken into account when calculating the radiated power and the final gravitational waveforms.}

Furthermore, we clarify the scaling when assuming a gravitating two-body system in virial equilibrium, i.e., when $r_g/r_{\rm orb}$ is of the order of $v_{\rm orb}^2$.
In this case, the angular frequency of the orbital motion is estimated as $\omega=v_{\rm orb}/r_{\rm orb}\sim v_{\rm orb}^3/r_g
$.
Therefore, there is the following hierarchy of the length scales~$r_g$, $r_{\rm orb}$, and $\omega^{-1} \sim \lambda_{\rm GW}$:
\begin{align}
    r_g\ll r_{\rm orb}\sim \frac{r_g}{v_{\rm orb}^2}\ll \frac{r_g}{v_{\rm orb}^3} \sim \frac{r_{\rm orb}}{v_{\rm orb}} \sim \omega^{-1}\sim \lambda_{\rm GW}\;, \label{v-scaling}
\end{align}
where $\lambda_{\rm GW}$ denotes the wavelength of the emitted GWs.\footnote{Thus, in the worldline EFT, each spatial derivative corresponds to 1PN order, while each temporal derivative corresponds to 1.5PN order. Since the effect of TLNs first enters at 5PN, the dTLNs (which are $\omega^2$ corrections to the TLNs) contribute at 8PN order.}

\subsection{Body zone description: perturbed compact object}\label{ssec:MST}
In the vicinity of a compact object, the tidal response can be formulated as a perturbation of the body's external metric.
Here, we focus on the dynamics of linear perturbations on the Schwarzschild spacetime in GR, whose line element is given by
\begin{align}\label{eq:bg_Sch}
    \de s^2=-f(r)\de t^2+\frac{\de r^2}{f(r)}+r^2(\de\theta^2+\sin^2\theta\,\de\varphi^2)\;,
\end{align}
where $f(r)\equiv 1-r_g/r$, with $r_g$ being the horizon radius.
Through a spherical harmonics decomposition of the metric perturbation, the linearized Einstein equations on the Schwarzschild background, expressed in the frequency domain, are reduced to ordinary differential equations for the radial part of each multipole.
Throughout this paper, we mainly focus on the odd-parity sector of the metric perturbations (and the corresponding gravito-magnetic response) for simplicity.
Note that for the Schwarzschild BH in GR, the even-parity sector is related to the odd-parity sector through the Chandrasekhar duality~\cite{Chandrasekhar:1975nkd,Chandrasekhar:1985kt,Solomon:2023ltn}, and therefore the gravito-electric tidal response should coincide with the gravito-magnetic tidal response~\cite{Hui:2020xxx,Katagiri:2023umb,Kobayashi:2025swn}.

The master equation for the odd-parity perturbations, known as the Regge-Wheeler equation~\cite{Regge:1957td}, takes the following form in the frequency domain:
\begin{align}
    \left[\left(f(r)\frac{\de}{\de r}\right)^2+\omega^2-f(r)V_{\rm RW}(r)\right]\Psi_{\rm RW}(r;\omega)=0\;. \label{GR-RW}
\end{align}
Here, $\Psi_{\rm RW}(r;\omega)$ denotes the master variable, which is constructed out of the odd-parity metric perturbations~\cite{Regge:1957td,Moncrief:1974am}.
The effective potential is given by
\begin{align}
    V_{\rm RW}(r)&=\frac{\ell(\ell+1)}{r^2}-\frac{3r_g}{r^3}\;,\label{GR-RW-potential}
\end{align}
with $\ell\,(\ge 2)$ being the multipole index.
Note that the magnetic number~$m$ does not appear explicitly due to the spherical symmetry of the background.

In the low-frequency regime ($r_g\omega \ll 1$), the analytic expression of homogeneous solutions of the Regge-Wheeler equation~\eqref{GR-RW} can be obtained using the MST formalism~\cite{Mano:1996gn,Mano:1996mf,Mano:1996vt} (see also Ref.~\cite{Sasaki:2003xr} for a review, and Ref.~\cite{Kobayashi:2025swn} for a brief summary).
To extract the external tidal perturbations and induced multipolar moments, we need the expressions of the horizon-ingoing solution of Eq.~\eqref{GR-RW} in the buffer zone ($r_g\ll r\ll r_{\rm orb}\ll \omega^{-1}$).
In this region, the general solution of the Regge-Wheeler equation~\eqref{GR-RW} can be expressed in the form of a series of hypergeometric functions~\cite{Mano:1996mf,Kobayashi:2025swn}.
Specifically, in terms of the dimensionless quantities~$x\equiv r/r_g$ and $\varepsilon\equiv r_g\omega$, the horizon-ingoing solution can be expressed as
\begin{align}
    \Psi_{\rm RW}=\Psi^{\text{hor-in}}_\nu(x)\equiv \Psi_\nu(x)+\Psi_{-\nu-1}(x)\;,\label{MST-hor-in-large-x}
\end{align}
with
\begin{align}
    \Psi_\nu(x)&=e^{-i\varepsilon x}\left(1-\frac{1}{x}\right)^{-i\varepsilon}x^{\nu+1}\sum_{n=-\infty}^\infty a_n^\nu \frac{\Gamma(-n-\nu-2-i\varepsilon)\Gamma(2n+2\nu+1)}{\Gamma(n+\nu+3-i\varepsilon)} \nonumber \\
    &\hspace{5mm}\times x^n {}_2F_1\left(-n-\nu-2-i\varepsilon,2-n-\nu-i\varepsilon;-2 n-2\nu;\frac{1}{x}\right) \nonumber \\
    &=C_\nu(\varepsilon)x^{\nu+1}\left[1+{\cal O}\left(x^{-1}, (\varepsilon x)^2\right)\right]\;, \label{growing}
\end{align}
and
\begin{align}
    \Psi_{-\nu-1}(x)
    &=e^{-i\varepsilon x}\left(1-\frac{1}{x}\right)^{-i\varepsilon}x^{-\nu}\sum_{n=-\infty}^\infty a_n^{\nu} \frac{\Gamma(n+\nu-1-i\varepsilon)\Gamma(-2n-2\nu-1)}{\Gamma(-n-\nu+2-i\varepsilon)} \nonumber \\
    &\hspace{5mm}\times x^{-n} {}_2F_1\left(n+\nu-1-i\varepsilon,3+n+\nu-i\varepsilon;2 n+2\nu+2;\frac{1}{x}\right) \nonumber \\
    &=C_{-\nu-1}(\varepsilon)x^{-\nu}\left[1+{\cal O}\left(x^{-1}, (\varepsilon x)^2\right)\right]\;, \label{decaying}
\end{align}
where in the second equality of Eqs.~(\ref{growing}) and (\ref{decaying}) we have introduced the coefficients $C_\nu(\varepsilon)$ and $C_{-\nu-1}(\varepsilon)$ for the growing and the decaying modes respectively in the limits $x \gg 1$ and $\varepsilon \ll 1$.
The coefficients~$a_n^\nu$ satisfy the following three-term recurrence relation:
\begin{align}
    \alpha_n^\nu a_{n+1}^\nu+\beta_n^\nu a_n^\nu+\gamma_n^\nu a_{n-1}^\nu=0\;, \label{3-term-reccurence}
\end{align}
with $a_0^\nu=1$ and
\begin{equation}
\begin{split}
    \alpha_n^\nu&=-i\varepsilon \frac{(n+\nu-1+i\varepsilon )(n+\nu-1-i\varepsilon )(n+\nu+1-i\varepsilon )}{(n+\nu+1)(2n+2\nu+3)}\;,\\
    \beta_n^\nu&=(n+\nu)(n+\nu+1)-\ell(\ell+1)+2\varepsilon ^2+\frac{\varepsilon ^2(\varepsilon ^2+4)}{(n+\nu)(n+\nu+1)}\;,\\
    \gamma_n^\nu&=i\varepsilon \frac{(n+\nu+2+i\varepsilon )(n+\nu+2-i\varepsilon )(n+\nu+i\varepsilon )}{(n+\nu)(2n+2\nu-1)}\;.
\end{split} \label{alpha-beta-gamma}
\end{equation}
Here, $\nu$ stands for the renormalized angular momentum, which is fixed by the convergence condition of the analytic series solution.
For a Schwarzschild background, $\nu$ is obtained as follows (see, e.g., Refs.~\cite{Mano:1996mf,Zhang:2013ksa,Casals:2015nja}):
\begin{equation}
\begin{split}
    \nu(\ell,\varepsilon)&\equiv\ell+\Delta\ell(\ell,\varepsilon)\;,\\
    \Delta\ell(\ell,\varepsilon)&=\sum_{n=1}^\infty\ell_{2n}\varepsilon^{2n}=-\frac{15 \ell^4+30 \ell^3+28 \ell^2+13 \ell+24}{2 \ell (\ell+1) (2 \ell-1) (2 \ell+1) (2 \ell+3)}\varepsilon^2+{\cal O}(\varepsilon^4)\;.
\end{split}\label{nu-expansion}
\end{equation}
For $\ell=2$, the leading coefficient of the renormalized angular momentum is given by $\ell_2=-\frac{107}{210}$.
This value agrees with the renormalized multipole moments originating from the ultraviolet divergence of a tail-of-tail diagram in the worldline EFT (see, e.g., Refs.~\cite{Goldberger:2009qd,Porto:2016pyg,Almeida:2021jyt,Ivanov:2025ozg}).
When $\nu$ is identified with the renormalized angular momentum, the set of coefficients~$\{a_n^\nu\}$ can be determined such that the formal solution given by Eqs.~\eqref{MST-hor-in-large-x}--\eqref{decaying} converges.
The low-frequency behavior of $a_n^\nu$ for several values of $n$ can be found in Refs.~\cite{Mano:1996mf,Casals:2015nja}.
It should be noted that if one is interested in the tidal response up to, e.g., ${\cal O}(\varepsilon^2)$, only the coefficients~$a_n^\nu$ with $|n|\le 2$ are relevant, as the others are of higher order in $\varepsilon$.

Before proceeding further, let us discuss some subtleties in the definition of the tidal response.

\paragraph{Problem of naive definition}
In the zero-frequency limit, the two independent modes~$\Psi_\nu$ and $\Psi_{-\nu-1}$ correspond to the tidal source and induced response field, respectively.
Therefore, one would naively define the bare tidal response as
\begin{align}
    \overline{\cal F}_\nu^{\rm B}(\omega)&=\frac{C_{-\nu-1}(\varepsilon)}{C_\nu(\varepsilon)}\;,\label{Bare-tidal-MST}
\end{align}
with the superscript~${\rm B}$ indicating the gravito-magnetic response.
However, there is an ambiguity in determining the amplitudes of the two modes, $C_{\nu}(\varepsilon)$ and $C_{-\nu-1}(\varepsilon)$.
Note that, as shown in Eqs.~\eqref{growing} and \eqref{decaying}, each of $\Psi_\nu$ and $\Psi_{-\nu-1}$ involves corrections of the form~$\left[1+{\cal O}(x^{-1},(\varepsilon x)^2)\right]$.
While the brackets are written with unit constant part, higher-order terms can generate additional $x^0$ contributions.
In particular, products such as $x^{-2}(\varepsilon x)^2$ yield constant terms, so the normalization of the constant part is contaminated by higher-order corrections.
As we will explain in Sec.~\ref{ssec:matching}, matching the MST solution to the worldline EFT gives one of the physically reasonable ways to resolve this ambiguity [see Eq.~\eqref{RW-metric-asymptotic-expansion}].
It should be noted that this subtlety does not affect the dTLNs at ${\cal O}(\varepsilon^2)$ for a Schwarzschild BH in GR, since the ${\cal O}(1)$ response vanishes, whereas the ${\cal O}(\varepsilon^3)$ response coefficient can be affected.

\paragraph{Problem of renormalization}
As pointed out in Refs.~\cite{Charalambous:2021mea,Perry:2024vwz,Kobayashi:2025swn}, the solutions~$\Psi_\nu$ and $\Psi_{-\nu-1}$ contain divergent terms when the limit~$\nu\to\ell$ is taken while treating the parameter~$\nu$ as independent of $\varepsilon$.
However, these divergent terms are unphysical since they cancel out when $\Psi_\nu$ and $\Psi_{-\nu-1}$ are summed up as in \eqref{MST-hor-in-large-x}.
In particular, the divergent coefficients of $C_{-\nu-1}$ appear as
\begin{align}
    C_{-\nu-1}\big|_{\nu=\ell+\delta\ell}
    =\frac{c_{\ell, 2}^{\rm div}\varepsilon^2+{\cal O}(\varepsilon^4)}{(-2\delta\ell)}+{\cal O}((\delta \ell)^0)\;,\label{pole}
\end{align}
with $c_{\ell, 2}^{\rm div}$ being a constant.
Here, the subscript~$\nu=\ell+\delta\ell$ indicates that we temporarily ignore the specific $\varepsilon$-dependence of $\delta\ell=\nu-\ell$ given in Eq.~\eqref{nu-expansion}.
In Ref.~\cite{Charalambous:2021mea}, it was pointed out that the pole part of $C_{-\nu-1}$ in \eqref{pole} can be interpreted as a counterterm to cancel the divergence in higher PN corrections to the source solution~$\Psi_\nu$.
These counterterms can source the renormalization flow of the tidal response coefficients, which should be interpreted as Wilsonian coupling coefficients in the worldline EFT~(see, e.g., Refs.~\cite{Porto:2016pyg,Levi:2018nxp,Ivanov:2022hlo,Saketh:2023bul,Mandal:2023hqa}).\footnote{The authors of Ref.~\cite{Saketh:2023bul} computed the renormalization group running of dTLNs up to ${\cal O}(\omega^2)$ for Kerr BHs.}
In Ref.~\cite{Charalambous:2021mea}, the renormalized tidal response function was introduced as
\begin{align}
    \overline{\cal F}_\ell^{\rm B,ren}(\omega)&=\left[\overline{\cal F}_{\nu=\ell+\delta\ell}^{\rm B}(\omega)\exp\left(-2\delta\ell\log x\right)\right]_{\text{subtract }(\delta\ell)^{-1}} \;,\label{reno}
\end{align}
where a logarithmic running part, $\left(c_{\ell, 2}^{\rm div}\varepsilon^2+{\cal O}(\varepsilon^4)\right)\log x$, remains even after the subtraction of the pole contribution.
Note in passing that the exponential factor in Eq.~\eqref{reno} corresponds to $x^{-2\delta\ell}$, which originates from the ratio~$x^{-\nu}/x^{\nu+1}$ with $\nu=\ell+\delta\ell$.
Note also that this reproduces the vanishing of the static TLNs in the $\varepsilon\to 0$ limit~\cite{Charalambous:2021mea,Kobayashi:2025swn}.\footnote{Although the absence of logarithmic terms at ${\cal O}(\varepsilon^0)$ may appear to be a fine-tuning from the EFT perspective~\cite{Porto:2016zng,Charalambous:2021mea,Charalambous:2022rre}, it can be explained by the hidden symmetry in the static limit of the worldline EFT (see e.g., Refs.~\cite{Ivanov:2022hlo,Parra-Martinez:2025bcu}).
Note, however, that this property no longer holds in beyond-GR setups~\cite{Barura:2024uog,Cano:2025zyk,Barbosa:2025uau,Bhattacharyya:2025slf}.}
In Sec.~\ref{ssec:matching}, we derive a similar renormalization equation by matching the MST results with those of the worldline EFT under dimensional regularization, where the spacetime dimension is taken to be $d=4-\epsilon$ and the multipole index is replaced by $\hat{\ell}=\ell/(d-3)$.
It should be noted that the finite part of the renormalized tidal response cannot be determined without fixing the renormalization scheme and the initial condition of the renormalization flow, while the logarithmic part is free from this ambiguity.
We will discuss this issue later in Sec.~\ref{sssec:ambiguity}.

\subsection{Post-Newtonian zone description: worldline picture}\label{ssec:worldline-EFT}
In the following, we review methods for incorporating finite-size effects into the PN zone description.
In this region, the binary constituent of interest is treated as a point particle, while the PN metric encodes the system's gravitational binding energy, tidal fields, and induced multipole moments.
Finite-size effects, such as multipole deformations~\cite{Goldberger:2005cd,Porto:2005ac,Goldberger:2007hy,Porto:2016pyg,Levi:2018nxp}, are captured by introducing internal degrees of freedom (DoFs) on the particle's worldline, which respond to the tidal field of the companion.
This approach allows one to study the long-distance dynamics without detailed knowledge of internal structure of the object.

\subsubsection{Worldline EFT}
Suppose the binary constituent of interest, approximated as a point particle, moves along a timelike worldline.
The corresponding EFT describing its localized DoFs coupled to gravity is called the worldline EFT~\cite{Goldberger:2004jt,Goldberger:2005cd,Goldberger:2006bd,Porto:2016pyg,Levi:2018nxp,Goldberger:2022ebt,Goldberger:2022rqf}.
The worldline is characterized by the center-of-mass position of the particle, $x_{\rm CM}^\mu(\tau)$, with $\tau$ being the proper time of the particle.
We shall use an orthonormal tetrad attached to the worldline, which consists of $e_0^\mu=u^\mu$, corresponding to the proper four-velocity of the particle's center of mass, and three unit spacelike vectors~$e_i^\mu$.\footnote{The tetrad specifies the local Lorentz frame, which is important in describing spinning point particles~(see, e.g., Refs.~\cite{Porto:2005ac,Porto:2007qi,Goldberger:2020fot}).}
In addition, we collectively denote the internal (hidden) DoFs by $X$, which, in the case of a regular star composed of fluids, correspond to the Lagrangian displacements of the fluid elements.
We can then introduce the multipole moments as composite operators~$Q_{L}^{\rm E/B}(X)$, which couple to gravito-electric/gravito-magnetic external tidal fields, respectively.
In this setup, the action of the worldline EFT takes the following form:
\begin{equation}
\begin{split}
    &S_{\rm WL}=S_{\rm pp}+S_X+S_{\rm int}\;,\\
    &S_{\rm pp}=-M\int\de\tau\;,\qquad
    S_X=\int\de\tau\;L_{X}(X,{\rm D}_\tau X, \cdots)\;,\\
    &S_{\rm int}=-\sum_{\ell\ge 2}\int \de\tau\; \left(Q^L_{\rm E}(X){\cal E}_L(\tau)+Q^L_{\rm B}(X){\cal B}_L(\tau)\right)\;,
\end{split}\label{worldline-action}
\end{equation}
where $M$ is the Arnowitt-Deser-Misner (ADM) mass of the object, $L_X$ is the Lagrangian for the internal DoFs~$X$, and ${\rm D}_\tau\equiv u^\mu\nabla_\mu$ denotes the derivative along the worldline, with $\nabla_\mu$ being the spacetime covariant derivative.
Also, the electric/magnetic tidal tensors~\cite{Thorne:1984mz,Zhang:1986cpa,Detweiler:2005kq,Taylor:2008xy,Poisson:2009qj,Poisson:2018qqd,Poisson:2020vap} are defined as
\begin{align}
\begin{split}
    {\cal E}_L(\tau)&\equiv \frac{1}{(\ell-2)!}{\rm D}_{\langle i_\ell}{\rm D}_{i_{\ell-1}}\cdots {\rm D}_{i_3}C_{|0|i_2|0|i_1\rangle}(x_{\rm CM}^\mu(\tau))\;, \\
    {\cal B}_L(\tau)&\equiv \frac{3}{(\ell-2)!(\ell+1)}{\rm D}_{\langle i_\ell}{\rm D}_{i_{\ell-1}}\cdots {\rm D}_{i_3}\star\!C_{|0|i_2|0|i_1\rangle}(x_{\rm CM}^\mu(\tau))\;,
\end{split}
\end{align}
with $\ell\ge 2$, where ${\rm D}_i\equiv e_i^\mu \nabla_\mu$, and we have defined
    \begin{align}
    C_{0i0j}=u^\mu e_i^\nu u^\rho e_j^\sigma C_{\mu\nu\rho\sigma}\;, \qquad
    \star C_{0i0j}=\frac{1}{2}u^\mu e_i^\nu u^\rho e_j^\sigma{\epsilon_{\mu\nu}}^{\alpha\lambda}C_{\alpha\lambda\rho\sigma}\;.
    \end{align}
Here, $C_{\mu\nu\rho\sigma}$ is the Weyl tensor, and $\epsilon_{\mu\nu\rho\sigma}$ is the completely anti-symmetric tensor associated with the bulk spacetime.

As mentioned earlier, in this paper, we mainly focus on the tidal response in the gravito-magnetic sector, 
while the formulation below can also be applied to the gravito-electric sector~(see, e.g., Refs.~\cite{Goldberger:2005cd,Ivanov:2022hlo}).\footnote{Note, however, that magnetic tidal effects arise at 1PN order higher than their electric counterparts, due to the velocity suppression of the background tidal tensor~\cite{Chia:2024bwc}.}
By evaluating the expectation value of $Q^L_{\rm B}$ in the presence of the background (classical) tidal tensor~$\overline{\cal B}_L$, we obtain the formula for linear response at the leading order of the interaction~\cite{Goldberger:2005cd}:
\begin{align}
    \langle{Q^L_{\rm B}(\tau)}\rangle_{\text{in-in}}&=-\int\de \tau'\; {G_{\rm ret,B}}^{L, L'}(\tau-\tau')\overline{\cal B}_{L'}(\tau')\;, \label{convolution}
\end{align}
where $Q^L_{\rm B}(\tau)\equiv Q^L_{\rm B}(X(\tau))$, and $\langle\,\rangle_{\text{in-in}}$ denotes the expectation value in the Schwinger-Keldysh (in-in) formalism~\cite{Schwinger:1960qe,Bakshi:1962dv,Bakshi:1963bn,Keldysh:1964ud,Jordan:1986ug} (see, e.g., Refs.~\cite{Chou:1984es,Calzetta_Hu_2008} for reviews and Refs.~\cite{Goldberger:2005cd,Galley:2009px} for applications to GW physics).
The retarded Green's function is given by
\begin{align}
    {G_{\rm ret,B}}^{L, L'}(\tau-\tau')&\equiv i\langle[Q^L_{\rm B}(\tau), Q^{L'}_{\rm B}(\tau')]\rangle_{\text{in-in}}\Theta(\tau-\tau')\;,\label{def-retarded-green-func}
\end{align}
where $[P,Q]\equiv PQ-QP$, and $\Theta$ denotes the Heaviside step function.
Note that the commutation relations are determined by the microscopic theory of the internal DoFs, which is described by $S_X$ in Eq.~\eqref{worldline-action}.
In the frequency domain, Eq.~\eqref{convolution} is expressed as
\begin{align}
     \langle{Q^L_{\rm B}(\omega)}\rangle_{\text{in-in}}&=-{G_{\rm ret,B}}^{L, L'}(\omega)\overline{\cal B}_{L'}(\omega)\;.\label{linear-response}
\end{align}
Here, we introduce a set of dimensionless tidal response functions~${\cal F}_\ell^{\rm B} (\omega)$ as follows:
\begin{align}
    {G_{\rm ret,B}}^{L, L'}(\omega)&\propto -r_g^{2\ell+1}{\cal F}_\ell^{\rm B}(\omega)\delta^{L,L'}\;,\label{retarded-Green's-function-freq}
\end{align}
where $\delta^{L,L'}=1$ if the multi-indices~$L$ and $L'$ contain the same set of indices (up to permutation), and $\delta^{L,L'}=0$ otherwise.
The overall scaling with $r_g^{2\ell+1}$ reflects the fact that the tidal response associated with the $\ell$-th multipole moment should scale as the $(2\ell+1)$-th power of the characteristic size of the body, in analogy with the Newtonian case.
Note also that there is no off-diagonal parts due to the spherical symmetry of the background~\cite{Charalambous:2021mea,Ivanov:2022hlo}.

Without specifying the microscopic theory of $Q^L_{\rm B}$, one cannot evaluate Eq.~\eqref{def-retarded-green-func}.
Nevertheless, certain general requirements constrain the expansion of the Green’s function.
First, the retarded Green's function should be analytic in the upper half of the complex $\omega$-plane due to causality (see, e.g., Ref.~\cite{Mizera:2023tfe}).
Also, assuming that all the poles or singularities are separated from the origin ($\omega =0$),\footnote{The poles of the retarded Green's function typically appear in the region~$|\omega|\sim r_g^{-1}$ with $\Im \omega\le 0$.}
the response function~${\cal F}_\ell^{\rm B}(\omega)$ can be expanded as a Taylor series around $\omega=0$.
Furthermore, since all physical quantities in the time domain are real-valued, the response function satisfies $({\cal F}_\ell^{\rm B})^* (\omega)={\cal F}_\ell^{\rm B}(-\omega)$, where $({\cal F}_\ell^{\rm B})^*$ denotes complex conjugation.
This allows us to write
\begin{equation}
\begin{split}
    \Re [{\cal F}_\ell^{\rm B}(\omega)]=&\frac{{\cal F}_\ell^{\rm B}(\omega)+{\cal F}_\ell^{\rm B}(-\omega)}{2}\;,\qquad
    \Im [{\cal F}_\ell^{\rm B}(\omega)]=\frac{{\cal F}_\ell^{\rm B}(\omega)-{\cal F}_\ell^{\rm B}(-\omega)}{2i}\;,
\end{split}    
\end{equation}
from which it follows that the real (imaginary) part of the tidal response function is an even (odd) function of $\omega$.
Therefore, we can expand the tidal response function as follows:
\begin{align}
    r_g^{2\ell+1}{\cal F}_\ell^{\rm B}(\omega)&=r_g^{2\ell+1}\left[\sum_{n\ge 0}{\cal K}_{\ell,(n)}^{\rm B}(r_g\omega)^{2n}+i \sum_{n\ge 0}{\cal N}_{\ell,(n)}^{\rm B}(r_g\omega)^{2n+1}\right]\;,\label{freq-resp}
\end{align}
where ${\cal K}_{\ell,(n)}^{\rm B}$ and ${\cal N}_{\ell,(n)}^{\rm B}$ are unspecified real-valued dimensionless coefficients.
Note in passing that we have kept the factor~$r_g^{2\ell+1}$ on both sides for later use in the renormalization analysis (see Sec.~\ref{sssec:renormalization}).
These coefficients correspond to the Wilsonian coupling coefficients when the internal DoFs are integrated out~\cite{Ivanov:2022hlo,Goldberger:2020fot}.
Note that the part even in $\omega$ corresponds to time-reversal invariant (or conservative) effects, while the odd part captures dissipative effects.\footnote{Even in Newtonian gravity, dissipative effects arise from the kinematic viscosity of the body~\cite{Poisson2014gravity}.
Viscosity leads to various dissipative phenomena, such as internal heating and the transfer of angular momentum between the body and the source, known as the ``tidal torque'' (see, e.g., Ref.~\cite{Thorne:1984mz}).}
Such time-asymmetric terms cannot be represented by a local action in the in-out formalism, reflecting the fact that dissipative interactions are inherently non-instantaneous (see Ref.~\cite{Ivanov:2022hlo}).

\subsubsection{Gravitational sector}
To complete the formalism, we have to take into account the gravitational action in the bulk spacetime, which we assume to be given by the Einstein-Hilbert action, i.e.,
\begin{align}
    S_{\rm bulk}=S_{\rm EH}\equiv \frac{1}{16\pi}\int\de^4x\sqrt{-g}\,R\;, \label{S_EH}
\end{align}
where $R$ is the Ricci scalar associated with the spacetime metric~$g_{\mu\nu}$.
In this paper, we assume that the exterior spacetime of the compact object is described by the static and spherically symmetric vacuum solution of GR, i.e., the Schwarzschild solution.
Since we focus on the region far from the object, the bulk metric can be expanded around the Minkowski background as follows:
\begin{align}
    g_{\mu\nu}
    =\eta_{\mu\nu}+h_{\mu\nu}\;.
\end{align}
In principle, one can go beyond the GR setup by including higher-dimensional operators or additional DoF(s), such as a scalar field (see, e.g., Refs.~\cite{Kuntz:2019zef,Bernard:2025dyh,Bhattacharyya:2025slf}).
We will comment on possible extensions in Sec.~\ref{sec:Extensions}.

\section{Matching between worldline EFT and BH perturbation}\label{sec:Matching}

In this section, we discuss the matching between the worldline EFT, describing the dynamics of a compact object in the PN zone, and BH perturbations in the body zone.
In the worldline picture, to describe the gravitational interaction between the binary constituents without explicitly resolving the microscopic DoFs~$X$, we use the in-in effective action~\cite{Goldberger:2020fot,Ivanov:2022hlo}, where the $X$-modes are integrated out along a closed-time-path contour (we follow the notation of Ref.~\cite{Galley:2009px}).
This approach is particularly suited for capturing the absorptive effects arising from the coupling between external perturbations and the hidden internal DoFs.
Once the $X$-modes are integrated out, the two-point function of $Q^L_{\rm B}$ (in the Hilbert space of $X$) becomes the Wilsonian coefficients, whose low-frequency expansion can be parametrized as in Eq.~\eqref{freq-resp}, without requiring the detailed information of the hidden sector.
In this section, we present the formulation for determining the bare values of the coefficients appearing in the low-frequency expansion by matching to the MST horizon-ingoing solution given in Eqs.~\eqref{MST-hor-in-large-x}--\eqref{decaying}.
We then discuss the renormalization flow and scheme dependence of the ${\cal O}(\omega^2)$ renormalized coefficients, which correspond to the dTLNs.

\subsection{Post-Newtonian zone calculation}\label{ssec:in-in}
\subsubsection{In-in effective action}
Let us consider the worldline description defined by the total action~$S=S_{\rm WL}+S_{\rm bulk}$, where $S_{\rm WL}$ is the worldline action given in Eq.~\eqref{worldline-action}, and $S_{\rm bulk}$ is the gravitational action in the bulk spacetime given in Eq.~\eqref{S_EH}.
To describe the dynamics including dissipative effects, we introduce the in-in effective action~$\Gamma^{\rm eff}_{\text{in-in}}$, which is defined through
\begin{equation}
    \exp\left(i\Gamma^{\rm eff}_{\text{in-in}}[{\bm{x}_{\rm CM}}_1,{h_{\mu\nu}}_1;{\bm{x}_{\rm CM}}_2,{h_{\mu\nu}}_2]\right) = \int \mathcal{D}X_1 \mathcal{D}X_2\; e^{iS[X_1,{\bm{x}_{\rm CM}}_1,{h_{\mu\nu}}_1]-iS[X_2,{\bm{x}_{\rm CM}}_2, {h_{\mu\nu}}_2]}\;,\label{def:in-in-action}
\end{equation}
where we have introduced two copies of each of $\bm{x}_{\rm CM}$, $h_{\mu\nu}$, and $X$, with the subscripts~$1$ and $2$ labeling the fields on the forward and backward branches of the closed-time-path contour, respectively.
Accordingly, we also introduce two copies of the magnetic tidal tensor~${\cal B}_L$ and the associated multipole moment operator~$Q^L_{\rm B}$.
The path integral is performed over the two copies of the hidden $X$ field(s).
All the fields satisfy the boundary conditions such that $X_1 = X_2$, ${\bm{x}_{\rm CM}}_1={\bm{x}_{\rm CM}}_2$, and ${h_{\mu\nu}}_1 = {h_{\mu\nu}}_2$ at the final time slice~$\tau=+\infty$.
Furthermore, we impose the initial boundary condition as $\rho_{\rm in}=\ket{0}\bra{0}\otimes \rho_X$, where $\ket{0}$ is the vacuum state of the graviton Fock space in Minkowski spacetime, and $\rho_X$ is the density operator representing the internal states.
Note that the Green's function for $Q^L_{\rm B}$ depends on the choice of $\rho_X$.\footnote{In fact, it was argued in Refs.~\cite{Goldberger:2019sya,Vidal:2024inh} that the appropriate boundary conditions, imposed either at the event horizon or at infinity, depend on the choice of the quantum vacuum in the BH background.
The classical absorptive boundary condition at the event horizon corresponds to the classical limit of the Boulware state~\cite{Goldberger:2019sya,Vidal:2024inh}.}
However, since we are concerned only with its low-frequency expansion, we do not specify $\rho_X$ explicitly.

To discuss the Green's function based on the in-in formalism, it is convenient to work in the Keldysh parametrization~\cite{Keldysh:1964ud}, 
\begin{equation}
    \begin{aligned}
    &X_- \equiv X_1 - X_2\;, \qquad X_+ \equiv \frac{1}{2}(X_1+X_2) \;,\\
    &{h_{\mu\nu}}_- \equiv {h_{\mu\nu}}_1 -{h_{\mu\nu}}_2\;, \qquad {h_{\mu\nu}}_+ \equiv \frac{1}{2} ({h_{\mu\nu}}_1 +{h_{\mu\nu}}_2) \;,\\
    &{\bm{x}_{\rm CM}}_- \equiv {\bm{x}_{\rm CM}}_1 - {\bm{x}_{\rm CM}}_2\;, \qquad {\bm{x}_{\rm CM}}_+ \equiv \frac{1}{2}({\bm{x}_{\rm CM}}_1+{\bm{x}_{\rm CM}}_2) \;,        
    \end{aligned}
\end{equation}
where $-/+$ variables are referred to as \textit{quantum/classical} variables.
We choose ${\bm{x}_{\rm CM}}_+=\bm{0}$, i.e., center-of-mass coordinates for the each time.
The $2\times 2$ Green's function for magnetic multipole moments $Q^L_{\rm B}$ is given by
\begin{equation}
    G_{AB} (\{Q^L_{\rm B}\})
    =
    \begin{pmatrix}
        G_{--} & G_{-+} \\
        G_{+-} & G_{++}
    \end{pmatrix}
    \equiv
    \begin{pmatrix}
        0 &  -i {G_{\rm adv,B}}^{L,L'}(\tau-\tau') \\
        -i {G_{\rm ret,B}}^{L,L'}(\tau-\tau') & \frac{1}{2} {G_{\rm H,B}}^{L,L'}(\tau-\tau')
    \end{pmatrix} \;,\label{Green'smatrix}
\end{equation}
where ${G_{\rm ret,B}}^{L,L'}$ denotes the retarded Green's function defined in Eq.~\eqref{def-retarded-green-func} and other components called the advanced and Hadamard Green's functions, respectively, are given by
\begin{equation}
\begin{split}
    {G_{\rm adv, B}}^{L,L'}(\tau-\tau')&=i\langle [Q^L_{\rm B}(\tau'),Q^{L'}_{\rm B}(\tau)]\rangle_{\text{in-in}}\Theta(\tau'-\tau)\;,\\
    {G_{\rm H,B}}^{L,L'}(\tau-\tau')&=\langle \{Q^L_{\rm B}(\tau), Q^{L'}_{\rm B}(\tau')\}\rangle_{\text{in-in}}\;,
\end{split}
\end{equation}
where $\{P,Q\}\equiv PQ+QP$.
The Feynman rules in the in-in formalism are very similar to those in the in-out formalism, but with contractions made over all closed-time-path indices~$A,B\in\{-,+\}$ with the following \textit{effective metric}:
\begin{equation}
    c^{AB} = c_{AB} = 
    \begin{pmatrix}
        0 & 1 \\
        1 & 0
    \end{pmatrix} \;.
\end{equation}

The in-in effective action introduced in Eq.~\eqref{def:in-in-action} can be decomposed as
\begin{align}
    \Gamma^{\rm eff}_{\text{in-in}}&=\Gamma^{\rm bulk}_{\text{in-in}}+\Gamma^{\rm pp}_{\text{in-in}}+\Gamma^{\rm finite}_{\text{in-in}}\;,
\end{align}
where the first two terms are $\Gamma^{\rm bulk}_{\text{in-in}}=S_{\rm bulk}[{h_{\mu\nu}}_1]-S_{\rm bulk}[{h_{\mu\nu}}_2]$ and $\Gamma^{\rm pp}_{\text{in-in}}=S_{\rm pp}[{\bm{x}_{\rm CM}}_1,{h_{\mu\nu}}_1]-S_{\rm pp}[{\bm{x}_{\rm CM}}_2,{h_{\mu\nu}}_2]$, respectively.
The last term, representing the finite-size effects, is given by
\begin{equation}
    \Gamma^{\rm finite}_{\text{in-in}}[{\bm{x}_{\rm CM}}_\pm,{h_{\mu\nu}}_\pm] = \frac{i}{2} \int\de\tau\,\de\tau' \langle {Q^{L}_{\rm B}}_A(\tau) {Q^{L'}_{\rm B}}_B(\tau') \rangle_{\text{in-in}} {{\cal B}_L}^A(\tau) {{\cal B}_{L'}}^B(\tau') \;,\label{BB-action}
\end{equation}
at leading order in the interaction between the internal DoFs and the tidal fields.

Note that the one-point function of $h_{\mu\nu}$ in the classical and static limit (i.e., in the absence of closed graviton loops and for $\omega\to 0$) reproduces the Schwarzschild metric in harmonic coordinates within GR~\cite{Duff:1973zz,Porto:2016pyg,Jakobsen:2020ksu,Mougiakakos:2020laz,Ivanov:2022hlo,Damgaard:2024fqj,Mougiakakos:2024nku}.
In general, the static and spherically symmetric BH solution of a given gravitational theory, described by the metric~$g_{\mu\nu}^{\rm BH}$, is expected to be reproduced from the one-point function of $h_{\mu\nu}$ in the corresponding worldline EFT as
\begin{align}
    g_{\mu\nu}^{\rm BH}=\eta_{\mu\nu}+\langle h_{\mu\nu}(\omega=0)\rangle_{\text{in-in}}^{\rm classical}\;.
\end{align}

\subsubsection{Metric perturbation with external tidal source}
In this section, we formulate the external tidal source and induced response terms based on the in-in effective action.
We extend the discussion in Refs.~\cite{Hui:2020xxx,Ivanov:2022hlo} to dynamical perturbations with characteristic frequency~$\omega=v_{\rm orb}/r_{\rm orb}\sim v_{\rm orb}^3/r_g$, assuming that the inspiralling system is in virial equilibrium.
Let us write the metric perturbation~$h_{\mu\nu}$ about the Minkowski background as
    \begin{align}
    h_{\mu\nu}=\overline{h}_{\mu\nu}+\delta h_{\mu\nu}\;.
    \end{align}
Here, $\overline{h}_{\mu\nu}$ corresponds to the external tidal perturbation, which is a homogeneous solution of the linearized Einstein equation.
Corresponding to the two copies of $h_{\mu\nu}$ (i.e., ${h_{\mu\nu}}_1$ and ${h_{\mu\nu}}_2$) introduced above, we also introduce copies for $\overline{h}_{\mu\nu}$ and $\delta h_{\mu\nu}$.
In what follows, we assume ${\overline{h}_{\mu\nu}}_1 = {\overline{h}_{\mu\nu}}_2 = {\overline{h}_{\mu\nu}}$, so that the quantum variable~${\overline{h}_{\mu\nu}}_-$ vanishes, while the classical variable becomes ${\overline{h}_{\mu\nu}}_+ = \overline{h}_{\mu\nu}$.
Moreover, we decompose $\delta h_{\mu\nu}$ as
\begin{align}
    \delta h_{\mu\nu}=\delta h_{\mu\nu}^{\rm pp}+\delta h_{\mu\nu}^{\rm finite}+\delta h_{\mu\nu}^{\rm finite,pp}\;,
\end{align}
where the superscripts~pp and finite indicate corrections due to point-particle interactions and finite-size effects, respectively.
For instance, the expectation value of $\delta h_{\mu\nu}^{\rm finite}$ can be calculated as follows:
\begin{align}
    \langle \delta h_{\mu\nu}^{\rm finite}(t,\bm{x}) \rangle_{\text{in-in}}^{\rm classical} &= \lim_{\hbar\to 0} \int \mathcal{D}\delta{h_{\alpha\beta}}_+ \mathcal{D}\delta{h_{\alpha\beta}}_- \left(\delta{h_{\mu\nu}}_+(t,\bm{x}) e^{i\Gamma^{\rm finite}_{\text{in-in}}+iS_{\rm bulk}[\delta{h_{\alpha\beta}}_1]-i S_{\rm bulk}[\delta{h_{\alpha\beta}}_2]-i\Gamma_{\rm GF}}\right) \;.\label{h-finite}
\end{align}
Here, we have included a gauge-fixing term~$\Gamma_{\rm GF}$ to impose the de Donder gauge~$\partial_\mu (h^{\mu\nu}-\frac{1}{2}\eta^{\mu\nu}h)=0$, which simplifies the calculation of Feynman diagrams.
Note that the classical limit~$\hbar\to 0$ is understood to be taken after temporarily restoring $\hbar$, even though we otherwise set $\hbar=1$.
In the classical limit, where graviton loops are ignored, one can circumvent the difficulties associated with the path integral over the metric and divergent counterterms possibly appearing in the bulk.

As mentioned earlier, we are focusing on the odd-parity sector, described by the Regge-Wheeler equation~\eqref{GR-RW}.
It is known that the following component of the Weyl tensor is related to the master variable~$\Psi_{\rm RW}$:
    \begin{align}
    C_{0rab}(t,\bm{x})&=2r^2\left[\partial_r (r^{-2}\nabla_{[a} h_{b]0})-\partial_0(r^{-2}\nabla_{[a} h_{b]r})\right] \nonumber \\
    &\propto \int\frac{\de\omega}{2\pi}e^{-i\omega t}\sum_{\ell\ge 2}\sum_{|m|\le\ell}\nabla_{[a}{Y^{({\rm T})}_{b]}}_{\ell m}r^{-1}\Psi_{\rm RW}(r;\omega)\;, \label{C0rab}
    \end{align}
where $\nabla_a$ denotes the covariant derivative on a unit two-sphere, and ${Y_a^{({\rm T})}}_{\ell m}(\theta, \varphi)$ denotes the transverse vector spherical harmonics.
Therefore, in what follows, we focus on this particular Weyl tensor component.
Corresponding to the metric perturbation, we introduce the decomposition of $C_{0rab}$ as follows:
    \begin{align}
    C_{0rab}=\overline{C}_{0rab}+\delta C_{0rab}^{\rm pp}+\delta C_{0rab}^{\rm finite}+\delta C_{0rab}^{\rm finite,pp}\;. \label{Weyl_decomposition}
    \end{align}

Let us first consider $\overline{C}_{0rab}$, which represents the external tidal perturbation.
As explained in the \hyperref[App:mathematical-formula]{Appendix}, this corresponds to a homogeneous solution of the linearized Einstein equation and can be expressed as [see Eq.~\eqref{Weyl-source-app}]
\begin{align}
    \overline{C}_{0rab}(t,\bm{x})&=\int\frac{\de\omega}{2\pi}\sum_{\ell\ge 2}\sum_{|m|\le\ell}c_{\ell m}^{\rm B}(\omega)e^{-i\omega t}j_\ell(\omega r)\nabla_{[a}{Y^{({\rm T})}_{b]}}_{\ell m}(\theta,\varphi) \nonumber \\
    &=\int \frac{\de\omega}{2\pi}\sum_{\ell\ge 2}\sum_{|m|\le\ell}\overline{c}_{\ell m}^{\rm B}(\omega)e^{-i\omega t}\left(\frac{r}{r_g}\right)^{\ell}\left[1+{\cal O}((\omega r)^2)\right]\nabla_{[a}{Y^{({\rm T})}_{b]}}_{\ell m}(\theta,\varphi)\;, \label{Weyl-source}
\end{align}
where $j_\ell$ denotes the spherical Bessel function, ${Y_{a}^{\rm (T)}}_{\ell m}$ denotes the transverse vector spherical harmonics, and $c_{\ell m}^{\rm B}$ are expansion coefficients depending on $\omega$.
In the second line, we have expanded the expression in powers of $\omega r$ and absorbed an overall constant factor into the coefficients~$\overline{c}_{\ell m}^{\rm B}$.
Equation~\eqref{Weyl-source} can be equivalently written in terms of the STF tensors (see, e.g., Refs.~\cite{Hui:2020xxx,Charalambous:2021mea,Iteanu:2024dvx}) as
\begin{align}
    \overline{C}_{0rab}(t,\bm{x})&= \epsilon_{ab}\int \frac{\de\omega}{2\pi}\sum_{\ell\ge 2} {\overline{c}^{\rm B}}_{L}(\omega )e^{-i\omega t}x^L\left[1+{\cal O}((\omega r)^2)\right]\;,\label{Weyl-source-STF}
\end{align}
where $\epsilon_{ab}$ is the completely anti-symmetric tensor on a unit two-sphere, and ${\overline{c}^{\rm B}}_{L}$ are the corresponding expansion coefficients.
Here, we recall that $x=r/r_g$.

Let us then consider the correction~$\delta C_{0rab}^{\rm finite}$, i.e., the contribution to $C_{0rab}$ arising from $\Gamma_{\text{in-in}}^{\rm finite}$.
This can be obtained from the path integral~\eqref{h-finite} by replacing the operator~$\delta{h_{\mu\nu}}_+$ with $\delta{C_{0rab}}_+$.
Following a calculation similar to Eq.~(2.31) of Ref.~\cite{Ivanov:2022hlo}, we obtain
\begin{equation}
\label{finite-positon}
    \begin{split}
        \langle \delta C_{0rab}^{\rm finite}(t,\bm{x}) \rangle_{\text{in-in}}^{\rm classical}
        & \approx
        i\int \de \tau \de \tau'\; \langle {\delta C_{0rab}}_+(t,\bm{x}){{\cal B}_L}_-(\bm{x}_{\rm CM}(\tau))\rangle_{\rm ret}\,
        {G_{\rm ret,B}}^{L,L'}(\tau-\tau') 
        \overline{\cal B}_{L'}(\bm{x}_{\rm CM}(\tau')) \;,
    \end{split}
\end{equation}
where $\langle\,\rangle_{\rm ret}$ denotes the retarded two-point function.
It should be noted that the magnetic tidal tensor is related to $C_{0rab}$ as
\begin{align}
    {\cal B}_L\propto {\rm D}_{\langle i_\ell}{\rm D}_{i_{\ell-1}}\cdots {\rm D}_{i_3}\delta_{i_2}^{a_2}\delta_{i_1\rangle}^{a_1}{\epsilon_{a_2}}^{rc}C_{0rca_1}\;, \label{B_C0rab}
\end{align}
with $\epsilon_{ijk}\equiv u^\mu\epsilon_{\mu ijk}$.
Also, the Weyl tensor is written in terms of the metric perturbation as in Eq.~\eqref{C0rab}, and therefore the retarded two-point function in Eq.~\eqref{finite-positon} can be written in terms of the retarded Green's function of the graviton,
\begin{align}
    {G_{\rm ret}^{\rm dD}}_{\mu\nu, \rho\sigma}(t,\bm{x};\tau,\bm{x}_{\rm CM}(\tau))
    =\left(\eta_{\mu(\rho}\eta_{\sigma)\nu}-\frac{1}{2}\eta_{\mu\nu}\eta_{\rho\sigma}\right)\Theta (t-\tau)\frac{\delta (t-\tau-|\bm{x}-\bm{x}_{\rm CM}(\tau)|)}{4\pi|\bm{x}-\bm{x}_{\rm CM}(\tau)|}\;. \label{Gret_graviton}
\end{align}
Note also that ${G_{\rm ret,B}}^{L,L'}(\tau-\tau')$ in Eq.~\eqref{finite-positon} encodes the information on the tidal response function [see Eq.~\eqref{retarded-Green's-function-freq}].

We will employ these equations in the next section, taking into account the point-particle interactions under dimensional regularization.

\subsubsection{Regularized calculation of tidal response function}\label{sssec:dim-reg}

Let us proceed to determine the gravito-magnetic tidal response function by matching the MST solution to the worldline EFT, extending the static-case analyses of Refs.~\cite{Hui:2020xxx,Ivanov:2022hlo} to a dynamical setup.
To regularize possible divergences arising in the momentum integration, we employ dimensional regularization, in which the spacetime dimension is analytically continued to a non-integer value, $d=4-\epsilon$, with $\epsilon$ being an infinitesimal parameter.
Accordingly, the multipole index~$\ell$ is replaced by $\hat{\ell}$, defined as
    \begin{align}
    \hat{\ell}\equiv \frac{\ell}{d-3}=\frac{\ell}{1-\epsilon}\;, \qquad
    \delta\ell\equiv \hat{\ell}-\ell=\ell\epsilon+{\cal O}(\epsilon^2)\;. \label{def_ell_hat}
    \end{align}
It is worth noting that the eigenvalues of the spherical harmonics on a unit $(d-2)$-dimensional sphere are given by $\ell(\ell+d-3)=(d-3)\hat{\ell}(\hat{\ell}+1)$.

First, we consider the contribution of the external tidal source to $C_{0rab}$, taking into account the point-particle interactions, which can be interpreted as PN corrections.
Specifically, we focus on the first two terms on the right-hand side of Eq.~\eqref{Weyl_decomposition}, where the expression for $\overline{C}_{0rab}$ is given by Eq.~\eqref{Weyl-source}.
The correction arising from the point-particle interactions, $\delta C_{0rab}^{\rm pp}$, can be represented diagrammatically, leading to
\begin{align}
    &\overline{C}_{0rab}(\bm{x};\omega)+\delta C_{0rab}^{\rm pp}(\bm{x};\omega)=\sum_{\ell\ge 2}\sum_{|m|\le\ell}c^{\rm B}_{\ell m}(\omega)\nabla_{[a}{Y_{b]}^{({\rm T})}}_{\ell m}(\theta,\varphi)\Bigg(j_{\hat{\ell}}(\omega r)+\raisebox{5pt}{\ExternalSource}\Bigg) \nonumber \\
    &=\sum_{\ell\ge 2}\sum_{|m|\le\ell}\overline{c}_{\ell m}^{\rm B}(\omega)\nabla_{[a}{Y_{b]}^{({\rm T})}}_{\ell m}(\theta,\varphi)\left(\frac{r}{r_g}\right)^{\hat{\ell}} \nonumber \\
    &\quad \times \left\{1+{\cal O}\left(\frac{r_g}{r}\right)+(\omega r)^2\left(a^+_{0, 1}+a^+_{1, 1}\frac{r_g}{r}+a^{+}_{2, 1}\left(\frac{r_g}{r}\right)^2+{\cal O}\left(\left(\frac{r_g}{r}\right)^3\right)\right)+{\cal O}((\omega r)^4)\right\} \nonumber \\
    &=\sum_{\ell\ge 2}\sum_{|m|\le\ell}\overline{c}_{\ell m}^{\rm B}(\omega)\nabla_{[a}{Y_{b]}^{({\rm T})}}_{\ell m}(\theta,\varphi)\left(\frac{r}{r_g}\right)^{\hat{\ell}} \nonumber \\
    &\quad \times\left\{1+{\cal O}\left(\frac{r_g}{r}\right)+(r_g\omega)^2\left(a^+_{0, 1}\left(\frac{r}{r_g}\right)^2+a^{+}_{1, 1}\frac{r}{r_g}+a^{+}_{2, 1}+{\cal O}\left(\frac{r_g}{r}\right)\right)+{\cal O}((r_g\omega)^4)\right\}\;, \label{point-particle}
\end{align}
where $a_{n,k}^+$, with $n,k \in \{0,1,2,\cdots\}$, denotes the coefficient of the correction term proportional to $(r_g/r)^n (\omega r)^{2k}$, which is a linear polynomial in $r_g \omega$.
In the Feynman diagram, a wavy line represents the off-shell graviton propagator (referred to as the potential mode) carrying four-momentum~$p_\mu$, with $p_0 \sim v/r$ and $|\bm{p}| \sim 1/r$ (see, e.g., Refs.~\cite{Goldberger:2004jt,Porto:2016pyg}).
The worldline is depicted by the vertical double line, and each vertex on it corresponds to the point-particle interaction.
Here, we ignore the graviton-loop structure within the shaded blob, as we work in the classical limit.
Note that the terms with the coefficients~$a_{0,k}^+$ simply originate from the expansion of the spherical Bessel function in powers of $\omega r$, whereas those with $a_{n,k}^+$ ($n\ge 1$) arise from diagrams containing $n$ graviton propagators.

Let us now consider the contribution from the finite-size effects to $C_{0rab}$, i.e., the last two terms on the right-hand side of Eq.~\eqref{Weyl_decomposition}.
Substituting Eqs.~\eqref{retarded-Green's-function-freq}, \eqref{B_C0rab}, and \eqref{Gret_graviton} into Eq.~\eqref{finite-positon}, performing dimensional regularization, and transforming to momentum space, we obtain
\begin{equation}
\begin{split}
    \langle \delta C_{0rab}^{\rm finite}(t,\bm{x}) \rangle_{\text{in-in}}^{\rm classical}
    \propto \epsilon_{ab}\sum_{\ell\ge 2}\int \frac{\de\omega}{2\pi}&\int\frac{\de^{D}\bm{p}}{(2\pi)^{D}}r_g^{2\ell+D-2}{\cal F}_{\hat{\ell}}^{\rm B}(\omega)\frac{(-i)^\ell p^L{\overline{c}^{\rm B}}_{L}(\omega)}{\bm{p}^2}e^{-i\omega t+i\bm{p}\cdot \bm{x}}\\
    &\times\left[1+{\cal O}((\omega r)^2)\right]\;,
\end{split}
\end{equation}
up to an overall numerical factor.
Using the formula for the momentum integral,
\begin{align}
    i^\ell\int \frac{\de^D\bm{p}}{(2\pi)^D} e^{i\bm{p}\cdot\bm{x}}\frac{p^L}{\bm{p}^2}&=\frac{\Gamma(D/2-1)\Gamma(2-D/2)}{2^\ell (4\pi)^{D/2}\Gamma(2-D/2-\ell)}x^L\left(\frac{\bm{x}^2}{4}\right)^{1-D/2-\ell}\;, \label{momentum-integ} 
\end{align}
we find
\begin{align}
    \langle \delta C_{0rab}^{\rm finite}(t,\bm{x}) \rangle_{\text{in-in}}^{\rm classical}
    =\sum_{\ell\ge 2}\sum_{|m|\le\ell}\int \frac{\de\omega}{2\pi}e^{-i\omega t}\overline{c}_{\ell m}^{\rm B}(\omega) r^{\hat{\ell}}\overline{\cal F}_{\hat{\ell}}^{\rm B}(\omega) \left(\frac{r_g}{r}\right)^{2\hat{\ell}+1}\left[1+{\cal O}((\omega r)^2)\right]
    \nabla_{[a}{Y^{({\rm T})}_{b]}}_{\ell m}(\theta,\varphi)\;, \label{finite-fourier}
\end{align}
where we have absorbed an overall numerical factor into $\overline{\cal F}_{\hat{\ell}}^{\rm B}(\omega)$.
Similarly to Eq.~\eqref{freq-resp}, we introduce the low-frequency expansion of $\overline{\cal F}_{\hat \ell}^{\rm B}$ as follows:
\begin{align}
    r_g^{2\hat{\ell}+1}\overline{\cal F}_{\hat{\ell}}^{\rm B}(\omega)&=r_g^{2\hat{\ell}+1}\left[\sum_{n\ge 0}\overline{\cal K}_{\hat{\ell},(n)}^{\rm B}(r_g\omega)^{2n}+i \sum_{n\ge 0}\overline{\cal N}_{\hat{\ell},(n)}^{\rm B}(r_g\omega)^{2n+1}\right]\;.\label{normalized-tidal-response}
\end{align}
Performing a Fourier transform of Eq.~\eqref{finite-fourier} and taking into account the point-particle interactions, we obtain
\begin{align}
    &\delta C_{0rab}^{\rm finite}(\bm{x};\omega)+\delta C_{0rab}^{\rm finite,pp}(\bm{x};\omega) \nonumber \\
    &=\sum_{\ell\ge 2}\sum_{|m|\le\ell}c^{\rm B}_{\ell m}(\omega)\nabla_{[a}{Y_{b]}^{({\rm T})}}_{\ell m}(\theta,\varphi)\Bigg(\raisebox{5pt}{\InducedResponse}+\raisebox{5pt}{\InducedResponseCorrected}\Bigg) \nonumber \\
    &=\sum_{\ell\ge 2}\sum_{|m|\le\ell}\overline{c}_{\ell m}^{\rm B}(\omega)\nabla_{[a}{Y_{b]}^{({\rm T})}}_{\ell m}(\theta,\varphi)\left(\frac{r}{r_g}\right)^{\hat{\ell}} \nonumber \\
    &\quad \times\overline{\cal F}_{\hat{\ell}}^{\rm B}(\omega)\left(\frac{r_g}{r}\right)^{2\hat{\ell}+1}\left\{1+{\cal O}\left(\frac{r_g}{r}\right)+(r_g\omega)^2\left(a^-_{0, 1}\left(\frac{r}{r_g}\right)^2+a^{-}_{1, 1}\frac{r}{r_g}+a^{-}_{2, 1}+{\cal O}\left(\frac{r_g}{r}\right)\right)+{\cal O}((r_g\omega)^4)\right\}\;. \label{worldline-response}
\end{align}
Combining Eqs.~\eqref{point-particle} and \eqref{worldline-response}, we finally find the following expression for the Weyl tensor component~$C_{0rab}$:
\begin{equation}
\begin{split}
    &C_{0rab}(\bm{x};\omega)
    =\overline{C}_{0rab}(\bm{x};\omega)+\delta C_{0rab}^{\rm pp}(\bm{x};\omega)+\delta C_{0rab}^{\rm finite}(\bm{x};\omega)+\delta C_{0rab}^{\rm finite,pp}(\bm{x};\omega) \\
    &=\;\sum_{\ell\ge 2}\sum_{|m|\le\ell}\overline{c}_{\ell m}^{\rm B}(\omega)\nabla_{[a}{Y_{b]}^{({\rm T})}}_{\ell m}(\theta,\varphi)\left(\frac{r}{r_g}\right)^{\hat{\ell}}\\
    &\times 
    \Bigg[1+{\cal O}\left(\frac{r_g}{r}\right)+(r_g\omega)^2\left(a^+_{0, 1}\left(\frac{r}{r_g}\right)^2+a^{+}_{1, 1}\frac{r}{r_g}+a^{+}_{2, 1}+{\cal O}\left(\frac{r_g}{r}\right)\right)+{\cal O}((r_g\omega)^4)\\
    &+\overline{\cal F}_{\hat{\ell}}^{\rm B}(\omega)\left(\frac{r_g}{r}\right)^{2\hat{\ell}+1}\left\{1+{\cal O}\left(\frac{r_g}{r}\right)+(r_g\omega)^2\left(a^-_{0, 1}\left(\frac{r}{r_g}\right)^2+a^{-}_{1, 1}\frac{r}{r_g}+a^{-}_{2, 1}+{\cal O}\left(\frac{r_g}{r}\right)\right)+{\cal O}((r_g\omega)^4)\right\}\Bigg]\;.
\end{split}\label{worldline-full-metric}
\end{equation}
Recall that we employ dimensional regularization with the spacetime dimension~$d = 4 - \epsilon$.
Accordingly, the coefficients~$a_{n,k}^\pm$ generally contain UV-divergent pieces proportional to $1/\epsilon$, in addition to finite parts (see, e.g., Refs.~\cite{Goldberger:2009qd,Porto:2016pyg,Almeida:2021jyt,Ivanov:2025ozg}).\footnote{See also Refs.~\cite{Ivanov:2024sds,Caron-Huot:2025tlq} for similar calculations of scalar Love numbers.}
These UV poles are canceled by the UV-divergent part of the bare tidal response function, leaving a finite renormalized response with logarithmic scale dependence, as we discuss in Sec.~\ref{sssec:renormalization}.

Since the Regge-Wheeler master variable~$\Psi_{\rm RW}$ is related to $C_{0rab}$ through Eq.~\eqref{C0rab}, we obtain the following expression in the buffer zone~$r_g\ll r\ll r_{\rm orb}\,(\ll \omega^{-1})$:
\begin{align}
    &\Psi_{\rm RW} 
    \propto \left(\frac{r}{r_g}\right)^{\hat{\ell}+1}\Bigg[1+{\cal O}\left(\frac{r_g}{r}\right)+(r_g\omega)^2\left(a^+_{0, 1}\left(\frac{r}{r_g}\right)^2+a^{+}_{1, 1}\frac{r}{r_g}+a^{+}_{2, 1}+{\cal O}\left(\frac{r_g}{r}\right)\right)+{\cal O}((r_g\omega)^4) \nonumber \\
    &+\overline{\cal F}_{\hat{\ell}}^{\rm B}(\omega)\left(\frac{r_g}{r}\right)^{2\hat{\ell}+1}\left\{1+{\cal O}\left(\frac{r_g}{r}\right)+(r_g\omega)^2\left(a^-_{0, 1}\left(\frac{r}{r_g}\right)^2+a^{-}_{1, 1}\frac{r}{r_g}+a^{-}_{2, 1}+{\cal O}\left(\frac{r_g}{r}\right)\right)+{\cal O}((r_g\omega)^4)\right\}\Bigg]\;, \label{RW-metric-asymptotic-expansion}
\end{align}
up to an overall constant factor, which will be matched to the MST solution given in Eqs.~\eqref{MST-hor-in-large-x}--\eqref{decaying}.

\subsection{Matching and ambiguities}\label{ssec:matching}
In this section, we determine the bare tidal response function in Eq.~\eqref{RW-metric-asymptotic-expansion} by matching it to the horizon-ingoing MST solution given in Eqs.~\eqref{MST-hor-in-large-x}--\eqref{decaying}.
The resulting bare response function contains pole terms in $\delta\ell$, which signal the need for renormalization.
In Sec.~\ref{sssec:renormalization}, we discuss the corresponding universal beta function, and in Sec.~\ref{sssec:ambiguity}, we examine the subtle ambiguities associated with the finite parts of the renormalized quantities.

\subsubsection{Matching in bare quantities}

As explained earlier, in the worldline EFT with dimensional regularization, the multipole index~$\ell$ is effectively shifted to $\hat{\ell}=\ell/(d-3)=\ell+\delta\ell$, which resolves the degeneracy between the two radial scalings~$r^{\ell+1}$ and $r^{-\ell}$. 
Meanwhile, in the MST formalism, an analogous role is played by the renormalized angular momentum~$\nu=\ell+\Delta\ell(\ell,\varepsilon)$. 
We therefore identify $\nu$ with $\hat{\ell}$, which allows the function~$\Psi_{\rm RW}$ in Eq.~\eqref{RW-metric-asymptotic-expansion} to be matched to the horizon-ingoing MST solution~\eqref{MST-hor-in-large-x} in the region where the regimes of validity of the two descriptions overlap.
Specifically, $\Psi_\nu(x)$ in Eq.~\eqref{growing} and $\Psi_{-\nu-1}(x)$ in Eq.~\eqref{decaying} correspond to the external tidal field and the induced response, respectively.
As a consequence, $\overline{\cal F}_\nu^{\rm B}(\omega)$ in Eq.~\eqref{Bare-tidal-MST} can be interpreted as the response function~$\overline{\cal F}_{\hat{\ell}}^{\rm B}(\omega)$ in the worldline EFT, which encodes the finite-size effect.
Written explicitly, we have
\begin{align}
    \overline{\cal F}_\nu^{{\rm B}}(\omega)
    &=\frac{\Gamma(-2\nu-1)\Gamma(\nu-1-i\varepsilon)}{\Gamma(-\nu+2-i\varepsilon)}\frac{\Gamma(\nu+3-i\varepsilon)}{\Gamma(2\nu+1)\Gamma(-\nu-2-i\varepsilon)} \nonumber \\
    &\quad \times\frac{1-i\varepsilon \frac{(\nu-1-i\varepsilon)(\nu+3-i\varepsilon)}{2\nu+2}+a_{-1}^\nu\left(-\frac{(-2\nu-1)(\nu+2-i\varepsilon)}{-\nu+2-i\varepsilon}\right)+{\cal O}(\varepsilon^2)}{1-i\varepsilon\frac{(-\nu-2-i\varepsilon)(-\nu+2-i\varepsilon)}{(-2\nu)}+a_{1}^\nu\left(-\frac{(2\nu+1)(-\nu+1-i\varepsilon)}{\nu+3-i\varepsilon}\right)+{\cal O}(\varepsilon^2)} \nonumber \\
    &\quad \times \frac{1+a_{2,1}^+\varepsilon^2+{\cal O}(\varepsilon^4)}{1+a_{2,1}^-\varepsilon^2+{\cal O}(\varepsilon^4)}\;.
    \label{full-F-nu}
\end{align}
Let us comment on the factor appearing in the third line.
The quantity~$\overline{\cal F}_\nu^{\rm B}(\omega)$ was defined in Eq.~\eqref{Bare-tidal-MST} as the ratio between the amplitudes of the two independent solutions, $\Psi_\nu$ and $\Psi_{-\nu-1}$.
However, as mentioned below Eq.~\eqref{Bare-tidal-MST}, the series expansions of the MST solutions contain an ambiguity in their constant terms, which makes the definition of $\overline{\cal F}_\nu^{\rm B}(\omega)$ subtle.
Matching to the asymptotic expression~\eqref{RW-metric-asymptotic-expansion} obtained in the worldline EFT resolves this ambiguity, and the factor in the third line of Eq.~\eqref{full-F-nu} arises from this matching.
Nevertheless, this factor is ignored in the following discussion, since it only affects terms of higher order in $\varepsilon$.\footnote{When we perform a double expansion with respect to $\varepsilon$ and $\delta\ell=\nu-\ell$, the first line of Eq.~\eqref{full-F-nu} already starts at first order in either $\varepsilon$ or $\delta\ell$. Therefore, the factor shown in the third line would only modify higher-order terms beyond the order retained in this section. Likewise, the ${\cal O}(\varepsilon^2)$ corrections appearing in the second line are neglected consistently.}

The response function in Eq.~\eqref{full-F-nu} can be expanded in terms of $\delta\ell=\nu-\ell$ as
\begin{align}
    &\overline{\cal F}_{\nu=\ell+\delta\ell}^{\rm B}(\omega)
    =\frac{[(\ell-2)!(\ell+2)!]^2}{(2\ell+1)!(2\ell)!}\Bigg[i\varepsilon+\frac{\varepsilon^2+{\cal O}(\varepsilon^4)}{(-2\delta\ell)}+\frac{\delta\ell}{2}\nonumber \\
    &\quad +\varepsilon^2\left(-\frac{7 \ell^4+14 \ell^3+3 \ell^2-4 \ell-2}{(\ell-1) \ell (\ell+1) (\ell+2) (2 \ell+1)}+2 {\rm Di} (2 \ell+1)-2 {\rm Di} (\ell-1)\right)+{\cal O}\left(\varepsilon^3, \varepsilon\delta\ell, (\delta\ell)^2\right)\Bigg]\;,
    \label{bare-matching}
\end{align}
where ${\rm Di}(x)=\frac{\de}{\de x}\log \Gamma(x)$ denotes the digamma function, and we have used the following formulae~\cite{Mano:1996mf,Casals:2015nja}:
    \begin{align}
    a^\nu_{1}=-i\varepsilon\frac{(\ell+3)^2}{2 (\ell+1) (2\ell+1)}+{\cal O}(\varepsilon^2,\delta\ell)\;, \qquad
    a^\nu_{-1}=-i\varepsilon\frac{(\ell-2)^2 }{2 \ell (2 \ell+1)}+{\cal O}(\varepsilon^2,\delta\ell)\;.
    \end{align}
The coefficient of $i\varepsilon$ in Eq.~\eqref{bare-matching} defines the TDNs, i.e.,
\begin{align}
    \overline{\cal N}_{\ell, (0)}^{\rm B}&=\frac{[(\ell-2)!(\ell+2)!]^2}{(2\ell+1)!(2\ell)!}\;,
\end{align}
where the difference from those obtained in, e.g., Refs.~\cite{Chakrabarti:2013lua,Charalambous:2021mea,HegadeKR:2024agt,Katagiri:2024wbg,Chakraborty:2025wvs} is due to different normalization factors.

\subsubsection{Renormalization}\label{sssec:renormalization}
The internal momentum integral in the worldline EFT exhibits both infrared (IR) and ultraviolet (UV) divergences.
The IR divergence arises from the long-range nature of the Newtonian potential, which can be absorbed into a redefinition of the time coordinate after resummation (see, e.g., Refs.~\cite{Goldberger:2009qd,Porto:2016pyg}).
In contrast, the UV divergence gives rise to the renormalization flow of the tidal response function.
Such divergent terms can be regularized using dimensional regularization, as introduced at the beginning of Sec.~\ref{sssec:dim-reg}.
Logarithmically UV divergent integrals yield a $1/\epsilon$ pole in the limit $\epsilon\to 0$, while power-law divergences can be set to zero.
These logarithmic divergences can then be handled following standard renormalization techniques in quantum field theory (see, e.g., Refs.~\cite{Porto:2016pyg,Levi:2018nxp} for reviews).

Let us now proceed to the renormalization of the tidal response function, following the procedure of Refs.~\cite{Porto:2016pyg,Levi:2018nxp,Mandal:2023hqa}.
Here, we assume that the $d$-dimensional Newton's constant~$G_d$, which has the mass dimension of $-d+2=-2+\epsilon$, scales as
\begin{align}
    G_d&\sim\mu^{\epsilon} G\;,
\end{align}
where 
$\mu$ is the renormalization energy scale and $G$ on the right-hand side is the four-dimensional Newton's constant.
Since the dimensionful bare tidal response function in Eq.~\eqref{freq-resp}, $r_g^{2\ell+1}{\cal F}_\ell^{\rm B}$, acquires an additional factor of $[\text{length}]^{2\delta\ell}$ under dimensional regularization, we adopt the following ansatz for the renormalization of the dimensionless tidal response function:
\begin{align}
    \overline{\cal F}_{\hat{\ell}}^{\rm B}&=\mu^{-2\delta\ell}\left(\overline{\cal F}_\ell^{\rm ren, B}(\mu;\omega)+\overline{\cal F}_\ell^{\rm ct, B}(\omega)\right)\;,\label{bare}
\end{align}
where the first term on the right-hand side represents the renormalized tidal response function, while the second term corresponds to the counterterm.
Note that the bare tidal response function on the left-hand side is independent of the renormalization scale~$\mu$.
Therefore, we obtain the renormalization flow equation as follows:
\begin{align}
    \mu\frac{\partial}{\partial\mu}\overline{\cal F}_\ell^{\rm ren, B}(\mu;\omega)&=-C_\ell^{\rm div, B}(\omega)+{\cal 
    O}(\epsilon)\;,\label{renormalization-flow}
\end{align}
where we have assumed $\overline{\cal F}_\ell^{\rm ct, B}(\omega)=-C_\ell^{\rm div, B}(\omega)/(2\delta\ell)+{\cal O}(\epsilon^0)$, with $\delta\ell=\ell\epsilon+{\cal O}(\epsilon^2)$.
From Eq.~\eqref{renormalization-flow}, the renormalized tidal response function can be obtained as
\begin{align}
    \overline{\cal F}_\ell^{\rm ren, B}(\mu;\omega)&=\overline{\cal F}_\ell^{\rm ren, B}(\mu_0;\omega)-C_\ell^{\rm div, B}(\omega)\log\left(\frac{\mu}{\mu_0}\right) \nonumber \\
    &=\overline{\cal F}_\ell^{\rm ren, B}(r_g^{-1};\omega)+C_\ell^{\rm div, B}(\omega)\log\left(\frac{r}{r_g}\right)\;, \label{renormalized-response}
\end{align}
where we have set the initial UV scale to be $\mu_0=r_g^{-1}$ and the energy scale relevant to the target system to be $\mu=r^{-1}$.\footnote{The logarithmic part remains universal so long as the perturbations outside the compact objects satisfy the Regge-Wheeler equation. 
Note, however, that the finite part of Eq.~\eqref{renormalized-response} is subject to ambiguities, as discussed in Sec.~\ref{sssec:ambiguity}.}

A caveat is in order. 
The authors of Ref.~\cite{Ivanov:2025ozg} showed that the structure of the renormalization flow described above is closely related to the anomalous dimensions of the multipole operators~$Q^L$. 
Specifically, they demonstrated that UV divergences from point-particle diagrams endow $Q^L$ with an anomalous scaling, and that this anomalous dimension is directly connected to the renormalized angular momentum~$\nu$ appearing in the MST formalism.\footnote{In particular, the effective multipole index~$\hat{\ell}$ associated with the anomalous dimension coincides with $\nu$. This is consistent with our procedure, in which $\nu$ is identified with $\hat{\ell}$.}
In this sense, the universal logarithmic running of the tidal response function, governed by the coefficient~$C_\ell^{\rm div,B}(\omega)$, is already encoded in the $\nu$-dependence of the MST solution. 
We emphasize, however, that our primary object of renormalization is not the multipole operator~$Q^L$ itself but the tidal response function, which directly captures the physical response of a compact object to external tidal fields.

In Eq.~\eqref{bare-matching}, the coefficient of the $1/\delta\ell$ pole corresponds to $C_\ell^{\rm div, B}$ introduced above, which gives rise to the logarithmic running of the tidal response function.
[Note in passing that the logarithmic running can also be inferred from Eq.~\eqref{reno}.]
Written explicitly, the logarithmic part of $\overline{\cal F}_\ell^{\rm ren, B}$ is given by
\begin{align}
    \overline{\cal F}_{\ell}^{\rm B,log}(\omega) 
    =\overline{\cal N}_{\ell,(0)}^{\rm B}\left[\varepsilon^2+{\cal O}(\varepsilon^4)\right]\log x\;, \label{log-coeff}
\end{align}
which reproduces the result in Refs.~\cite{Katagiri:2024wbg,Chakraborty:2025wvs}.
The presence of the logarithmic running in dTLNs suggests the absence of hidden symmetry in the Regge-Wheeler equation at finite frequency, in contrast to the case of static perturbations~\cite{Porto:2016zng,Charalambous:2021kcz,Hui:2021vcv,BenAchour:2022uqo,Hui:2022vbh,Charalambous:2022rre,Katagiri:2022vyz,Berens:2022ebl,Sharma:2024hlz}.
Due to the logarithmic running, the dTLNs cannot be set to zero at all distance scales; even if they vanish at one scale, they reemerge at other scales through renormalization.
Consequently, the dTLNs leave a physical imprint on the inspiral waveform, appearing from the 8PN order.

\subsubsection{Ambiguities of the finite part}\label{sssec:ambiguity}
The finite part of Eq.~\eqref{renormalized-response} remains undetermined because it depends on the particular scheme used to separate the counterterms in Eq.~\eqref{bare}.
This scheme dependence introduces ambiguities when one attempts to define the renormalized tidal response function and, consequently, the corresponding dTLNs.

To avoid such ambiguities, one can instead compute the inspiral gravitational waveform using the bare tidal response function.
In this approach, the tidal response is evaluated by analytically continuing the multipole index~$\ell$, and the waveform is obtained without explicitly performing the renormalization.
As a result, the PN expansion of the waveform does not contain the logarithmic correction term proportional to $\log v$, which would otherwise arise from the renormalization-scale dependence of the tidal response~\cite{Chakraborty:2025wvs}.
Also, the PN expansion acquires non-integer powers, enabling a clear separation between the point-particle and finite-size contributions~\cite{Creci:2021rkz}.
It should be noted, however, that this decomposition involves divergent pieces that cancel among themselves in the final physical result.

Another way to avoid the aforementioned ambiguities is to compare each coefficient of the tidal response function within a specified renormalization scheme or under a fixed renormalization condition.
Since the initial value of the renormalization flow at the UV scale ($\mu_0=r_g^{-1}$) cannot be specified a priori, we adopt the minimal subtraction (MS) scheme for the bare tidal response function in Eq.~\eqref{bare-matching}, which is determined by matching to the MST solution.
In summary, our definition of the renormalized dTLNs relies on i) the matching between the MST solution and the worldline-EFT solution and ii) the renormalization with the MS scheme.

Let us now extract the dTLNs from the renormalized tidal response function, adopting the MS scheme.
Subtracting the terms of ${\cal O}(1/\delta\ell)$ and replacing $\delta\ell$ with $\Delta\ell$ (i.e., the shift of the multipole index associated with the renormalized angular momentum) in Eq.~\eqref{bare-matching},
the finite part of Eq.~\eqref{renormalized-response} is obtained as
\begin{align}
    \overline{{\cal F}}_\ell^{\rm ren, B}(r_g^{-1};\omega)
    =\overline{\cal N}_{\ell, (0)}^{\rm B}\Bigg[&i\varepsilon+\varepsilon^2\left(-\frac{7 \ell^4+14 \ell^3+3 \ell^2-4 \ell-2}{(\ell-1) \ell (\ell+1) (\ell+2) (2 \ell+1)}+2 {\rm Di} (2 \ell+1)-2 {\rm Di} (\ell-1)\right) \nonumber \\
    &+\frac{\Delta\ell(\ell,\varepsilon)}{2}+{\cal O}(\varepsilon^3)\Bigg]\;. \label{F^ren_MS}
\end{align}
In the zero-frequency limit ($\varepsilon\to 0$), 
the tidal response function vanishes for all $\ell\ge 2$, corresponding to the vanishing of the static TLNs.
The coefficient of $i\varepsilon$, namely the TDNs~$\overline{\cal N}_{\ell, (0)}^{\rm B}$, gives rise to the leading dissipative effect, which is free from logarithmic running~\cite{Kobayashi:2025swn}.
Substituting the expression~\eqref{nu-expansion} for $\Delta\ell$ into Eq.~\eqref{F^ren_MS} and adding the logarithmic part from Eq.~(\ref{log-coeff}), the dTLNs of ${\cal O}(\varepsilon^2)$ can be read off as
\begin{align}
    \overline{\cal K}_{\ell, (1)}^{\rm B}=\;\overline{\cal N}_{\ell, (0)}^{\rm B}
    \Bigg[&-\frac{127 \ell^6+381 \ell^5+216 \ell^4-203 \ell^3-151 \ell^2+14 \ell-24}{4 (\ell-1) \ell (\ell+1) (\ell+2) (2 \ell-1) (2 \ell+1) (2 \ell+3)} \nonumber \\
    &+2{\rm Di} (2 \ell+1) - 2{\rm Di} (\ell-1)+\log x\Bigg]\;, \label{dTLN}
\end{align}
which is non-vanishing and exhibits logarithmic running.
For example, we obtain
    \begin{align}
    \overline{\cal K}_{\ell=2, (1)}^{\rm B}=\frac{1}{5}\left(\frac{71}{35}+\log x\right)\;, \quad
    \overline{\cal K}_{\ell=3, (1)}^{\rm B}=\frac{1}{252}\left(\frac{337}{210}+\log x\right)\;, \quad
    \overline{\cal K}_{\ell=4, (1)}^{\rm B}=\frac{1}{7056}\left(\frac{20561}{13860}+\log x\right)\;.
    \end{align}
Note that the finite part differs from those in Refs.~\cite{HegadeKR:2024agt,Katagiri:2024wbg,Chakraborty:2025wvs} due to the aforementioned scheme dependence. 
However, this ambiguity in the choice of renormalization scheme does not affect physical observables, such as the binding energy of the two-body system or the final emitted gravitational waveform---it merely reflects a different parametrization of the same phenomenon.

\section{Possible extensions}\label{sec:Extensions}
In this section, we discuss possible extensions of our formalism beyond BH solutions in vacuum GR.
In Sec.~\ref{ssec:Compact}, we will consider general static and spherically symmetric compact objects whose external spacetime is described by the vacuum solution in GR.
In Sec.~\ref{ssec:parametrized}, we will consider cases where the dynamics of metric perturbations is governed by a master equation that is slightly modified from the Regge-Wheeler equation in GR, applicable to BHs in modified gravity as well as to BHs surrounded by matter fields.

\subsection{Generic non-rotating compact objects in GR}\label{ssec:Compact}

First, we discuss the dynamical tidal response of a generic compact object whose exterior spacetime is given by the Schwarzschild solution in GR.
Let the object have ADM mass~$M$ and radius~${\cal R}$, with ${\cal R}$ assumed to be larger than the Schwarzschild radius~$r_g = 2M$.
Equivalently, in terms of the compactness~${\cal C}\equiv M/{\cal R}$, we assume ${\cal C}<1/2$.
Moreover, we assume that the separation of scales discussed in Sec.~\ref{ssec:scale} also applies here, with $r_g$ therein replaced by ${\cal R}$.

As in the previous sections, we focus on odd-parity metric perturbations.
Therefore, the perturbations outside the object are governed by the Regge-Wheeler equation~\eqref{GR-RW}, and the solution can be expressed as a linear combination of $\Psi_\nu$ and $\Psi_{-\nu-1}$ [given in Eqs.~\eqref{growing} and \eqref{decaying}, respectively] as
\begin{align}
    \Psi_{\rm RW}(r;\omega)&=
    \Psi_\nu(x)+{\cal T}_{-\nu-1}\Psi_{-\nu-1}(x)\;, \label{RW_ext_sol}
\end{align}
with $x=r/r_g$.
Here, the coefficient~${\cal T}_{-\nu-1}$ is determined from the matching condition at the surface of the compact object.
For this purpose, it is convenient to define the following quantity~\cite{Kojima:1992ie,Saketh:2024juq}:
\begin{align}
    T(\omega)&=\left.\frac{r}{r_*}\frac{\de \log \Psi_{\rm RW}}{\de\log r_*}\right|_{r={\cal R}}\;,\label{matching-CO}
\end{align}
with $r_*$ being the tortoise coordinate, which should take the same value when evaluated on both sides of the object's surface.
Given the equation of state of the object, the interior solution can be obtained by solving equations governing the coupling between gravity and matter. 
Then, imposing the continuity conditions on both the field and its first derivative at $r = {\cal R}$ allows us to connect the information of the interior structure to that of the exterior regime. 
Practically, one expands $T(\omega)$ in powers of ${\cal R}\omega\,(\ll 1)$ as 
    \begin{align}
    T(\omega)=T_0-i{\cal R}\omega T_1-({\cal R}\omega)^2T_2+\cdots\;,
    \end{align} 
where all the coefficients $T_0$, $T_1$, $T_2$, $\cdots$ capture the interior information. 
Once these coefficients are determined to the required order, the coefficient~${\cal T}_{-\nu-1}$ in Eq.~\eqref{RW_ext_sol} can be obtained from the continuity conditions.
(For instance, to compute static TLNs and TDNs, one only needs $T_0$ and $T_1$, while $T_2$ is required to determine dTLNs.)
Note that the functional forms of $\Psi_\nu(x)$ and~$\Psi_{-\nu-1}(x)$ are specified by $\varepsilon = r_g \omega$, or equivalently by ${\cal R}\omega$, given that the compactness~${\cal C}$ is fixed.
Consequently, the coefficient~${\cal T}_{-\nu-1}$ is a function of ${\cal R}\omega$, the compactness~${\cal C}$, and the parameters~$T_0, T_1, T_2, \cdots$.
Also, since the structure of the worldline EFT remains unchanged from the case of BHs in GR, the matching to the MST solution proceeds in the same way, with the UV scale for renormalization taken as $\mu_0={\cal R}^{-1}$.
With these notions, the bare tidal response function is given as follows:
\begin{align}
    \overline{\cal F}_\nu^{\rm B}({\cal R}\omega, {\cal C}, T_0, T_1, T_2, \cdots)&={\cal T}_{-\nu-1}\overline{\cal F}_\nu^{\rm B\,(BH)}\;,\label{bare-CO}
\end{align}
where $\overline{\cal F}_\nu^{\rm B\,(BH)}$ denotes the tidal response function of a Schwarzschild BH in GR, with the same ADM mass.\footnote{The authors of Ref.~\cite{Saketh:2024juq} proposed subtracting the total scattering amplitude of a Schwarzschild BH from that of a neutron star of the same mass. This procedure cancels the contributions from GW scattering off the background metric, isolating the difference arising from their tidal responses.}
Notice that with this bare tidal response function it is possible to directly obtain physical observables such as the gravitational waveforms without using any specific renormalization scheme. 

\subsection{Beyond GR and environmental effects}\label{ssec:parametrized}

In this section, we consider a theory-agnostic approach to describe perturbations about BHs beyond the GR vacuum background, which apply to both BHs in modified gravity theories and BHs immersed in a matter environment.
The assumptions we impose hereafter are (i)~that the background geometry is described by an asymptotically flat, static, spherically symmetric BH~\footnote{The staticity and spherical symmetry of the background can be relaxed if one, for instance, considers the modified Teukolsky equations within a stationary and axisymmetric spacetime background (see, e.g., Refs.~\cite{Li:2022pcy,Hussain:2022ins,Cano:2023tmv,Cano:2023jbk,Wagle:2023fwl,Cano:2024jkd,Cano:2025zyk}).} with a (non-degenerate) horizon at $r=r_g$, (ii)~that deviations from vacuum GR are small, and (iii)~the absence of couplings to other physical DoFs.
With these assumptions, we focus on odd-parity metric perturbations and consider a deformation of the Regge-Wheeler equation of the following form:
\begin{align}
   \left[\left(F\frac{\de}{\de x}\right)^2+\varepsilon^2-F\overline{V}_\ell\right]\Psi_{\rm dRW}=0\;,\label{gRW-canonical}
\end{align}
where $x\equiv r/r_g$, $\varepsilon\equiv r_g\omega$, and $F(x)\equiv f(x)Z(x)$, with $f(x)\equiv 1-1/x$ and $Z(x)$ assumed to be regular at $x=1$.
The function $Z(x)$ represents the deformation from the Regge-Wheeler equation; Eq.~\eqref{gRW-canonical} with $Z(x) = 1$ reduces to the Regge-Wheeler equation.  
Here, $\overline{V}_\ell$ is the effective potential, satisfying $\overline{V}_\ell(x)\to r_g^2V_{\rm RW}(r_gx)$ in the limit of vacuum GR, with $V_{\rm RW}$ being the Regge-Wheeler potential defined in Eq.~\eqref{GR-RW-potential}.
Also, $\Psi_{\rm dRW}$ denotes the master variable of the deformed Regge-Wheeler equation, which we redefine as $\psi_{\rm dRW}=\sqrt{Z}\,\Psi_{\rm dRW}$ in order to rewrite Eq.~\eqref{gRW-canonical} in the following form:
\begin{equation}
   \left[\left( f\frac{\de}{\de x}\right)^2+\frac{\varepsilon^2}{Z^2} -f\left(r_g^2V_{\rm RW}+\delta V_\ell\right)\right]\psi_{\rm dRW}=0\;, 
   \label{deformed-RWeq}
\end{equation}
where $\delta V_\ell$ is given by
    \begin{align}
    \delta V_\ell(x)=\frac{\overline{V}_\ell}{Z}+\frac{1}{\sqrt{Z}}\frac{\de}{\de x}\left( f\frac{\de\sqrt{Z}}{\de x}\right)-r_g^2V_{\rm RW}(r_gx)\;.\label{deformed-RW}
    \end{align}
In the vacuum GR case where $Z=1$ and $\overline{V}_\ell(x)=r_g^2V_{\rm RW}(r_gx)$, one has $\delta V_\ell=0$.
Note in passing that, in the absence of couplings to other physical DoFs, the master equation for even-parity metric perturbations can also be recast in the form~\eqref{deformed-RWeq} by use of the Chandrasekhar transformation (see, e.g., Ref.~\cite{Katagiri:2023umb}).

Let us assume that the deviations from vacuum GR can be treated perturbatively, and write $Z(x)=1+\delta Z(x)$.
Then, to leading order in $\delta Z$, Eq.~\eqref{deformed-RWeq} can be rewritten in the following form~\cite{Cardoso:2019mqo}:
    \begin{align}
    \left[\left( f\frac{\de}{\de x}\right)^2+\tilde{\varepsilon}^2-f\left(r_g^2V_{\rm RW}+\delta \tilde{V}_\ell\right)\right]\psi_{\rm dRW}=0\;, \label{deformed-RWeq2}
    \end{align}
where we have defined
    \begin{align}
    \tilde{\varepsilon}^2\equiv \varepsilon^2(1-2\delta Z_1)\;, \qquad
    \delta\tilde{V}_\ell\equiv \delta V_\ell+\frac{2\varepsilon^2}{f}(\delta Z-\delta Z_1)\;,
    \end{align}
with $\delta Z_1\equiv \delta Z(x=1)$.
Note that $\delta\tilde{V}_\ell$ is regular at $x=1$.
Written in this form, it is straightforward to obtain a perturbative solution for $\psi_{\rm dRW}$.
At linear order in the deviations from vacuum GR, Eq.~\eqref{deformed-RWeq2} yields the following differential equation for $\delta\psi\equiv\psi_{\rm dRW}-\Psi_{\rm RW}(r_gx;r_g\tilde{\varepsilon})$:
    \begin{align}
    \left[\left( f\frac{\de}{\de x}\right)^2+\varepsilon^2-r_g^2fV_{\rm RW}\right]\delta\psi
    =f\Psi_{\rm RW}(r_gx;r_g\varepsilon)\delta\tilde{V}_\ell\;,
    \end{align}
which is subject to the ingoing boundary condition at the horizon and can be solved using the Green's function method.
Recall that $\Psi_{\rm RW}(r;\omega)$ is the homogeneous solution to the Regge-Wheeler equation~\eqref{GR-RW} in vacuum GR.
One often assumes the correction to the effective potential, $\delta\tilde{V}_\ell$, to take the form of a series expansion in $r_g/r$, in order to capture modified-gravity or environmental effects in a model-independent way. 
This approach, known as the parametrized formalism, has been widely used to study quasinormal modes \cite{Cardoso:2019mqo,McManus:2019ulj}, static TLNs \cite{Katagiri:2023umb}, and TDNs of BHs~\cite{Kobayashi:2025swn}.

In this setup, however, it becomes necessary to extend the worldline EFT in order to determine the frequency-dependent tidal response, particularly beyond the ${\cal O}(\varepsilon^2)$ terms.
In particular, one needs to include additional DoF(s) or higher-curvature corrections in the bulk spacetime.
These modifications introduce new types of interaction vertices in the bulk and, consequently, affect the PN corrections to the tidal response function (see, e.g., Refs.~\cite{Endlich:2017tqa,Kuntz:2019zef,Wong:2019yoc,Bernard:2023eul,Bhattacharyya:2023kbh,Almeida:2024cqz,Bhattacharyya:2025slf,Wilson-Gerow:2025xhr,Bernard:2025dyh}).\footnote{See also, e.g., Ref.~\cite{Modrekiladze:2024htc} for a worldline EFT description of viscous-fluid environments based on the Schwinger-Keldysh approach.}
Moreover, non-trivial background configurations of the additional fields spontaneously break part of the diffeomorphism invariance by selecting a preferred slicing of spacetime (see Refs.~\cite{Franciolini:2018uyq,Mukohyama:2022enj,Mukohyama:2022skk,Mukohyama:2023xyf,Aoki:2023bmz,Tomizuka:2025dpy,Mukohyama:2025jzk} and references therein).
This allows one to construct the most general action consistent with the unbroken symmetries.
Within the worldline EFT for such symmetry-broken theories, the propagation speed (phase velocity) of gravitons may deviate from unity, which alters the ${\cal O}(\varepsilon^2)$ terms in the tidal response.
Hence, the dynamical tidal corrections must be computed within a specific worldline EFT that accurately describes the chosen setup.
Performing such matching calculations provides a concrete connection between the model parameters and the observables in the gravitational waveform.
We leave a detailed investigation of these extensions to future work.

\section{Conclusions}\label{sec:Conclusions}

We have reconsidered the relativistic formulation of the tidal response of non-rotating black holes (BHs) within general relativity (GR), applicable to dynamical perturbations with non-zero frequency.
Our formulation is based on the matched asymptotic expansion method, which exploits the separation of scales during the inspiral phase (see Sec.~\ref{ssec:scale}).
This separation ensures the validity of the post-Newtonian (PN) expansion for modeling the inspiral waveform.
One can, in principle, determine the bare tidal response function to all orders in frequency by matching the horizon-ingoing Mano-Suzuki-Takasugi (MST) solution with the worldline effective field theory (EFT) result obtained in dimensional regularization.
In this work, we have focused on the tidal response coefficients of the frequency-squared term, known as the dynamical tidal Love numbers (dTLNs), and found that the renormalization flow of the tidal response function induces the logarithmic running of the dTLNs (see Sec.~\ref{ssec:matching}).
These renormalization flows originate from the ultraviolet divergences in the metric sourced by the point-particle action [see Eq.~\eqref{renormalization-flow}], while they are already incorporated in the MST solution through the renormalized angular momentum~$\nu$ defined in Eq.~\eqref{nu-expansion}.
To avoid ambiguities in the finite part of the tidal response function, we have adopted the minimal subtraction scheme in Sec.~\ref{sssec:ambiguity} to obtain the formula~\eqref{dTLN} for the dTLNs.
The non-vanishing and logarithmically running dTLNs indicate that deviations from point-particle behavior arise even in the conservative process within GR, appearing at 8PN order.
In summary, our formulation provides a consistent and systematic approach to computing the dynamical tidal response of non-rotating BHs, capturing its renormalization and PN behavior within GR.

Several promising directions remain to be explored in future work.
As discussed in Sec.~\ref{sec:Extensions}, it would be intriguing to extend our framework to more general settings, including neutron stars, exotic compact objects, and BHs in modified gravity theories or those embedded in environmental fields.
It is also interesting to extend our formulation to rotating backgrounds. 
Another direction is to compute the inspiral gravitational waveform directly from the bare tidal response function, thereby avoiding ambiguities associated with renormalization.
It would also be worthwhile to assess the observational prospects for detecting the effects of the dynamical tidal response.
Pursuing these directions will help bridge the gap between the fundamental theory of gravity and the parameters characterizing the gravitational-wave signals observed during the inspiral phase.
We leave these investigations for future work.

\paragraph{Note added}
While we were finalizing this project and submitting v1 of this manuscript on arXiv, we became aware of Ref.~\cite{Combaluzier--Szteinsznaider:2025eoc} by Oscar Combaluzier-Szteinsznaider, Daniel Glazer, Austin Joyce, Maria J.~Rodriguez, and Luca Santoni, which addresses a similar problem.
In contrast to their work, our approach employs the MST formalism explicitly, in which the renormalized angular momentum makes the separation between the tidal source and the induced response manifest.
We also clarified that the logarithmic running in the dTLNs, relevant to the ultraviolet divergences in the worldline EFT, can be captured by the renormalized angular momentum in the MST method, already in v1 of this manuscript.
After submitting v1 of this manuscript on arXiv, in light of Ref.~\cite{Combaluzier--Szteinsznaider:2025eoc}, we partly refined the presentation of our formalism in Secs.~\ref{ssec:worldline-EFT} and \ref{ssec:in-in} by employing the proper time along the worldline.
We thank the authors of Ref.~\cite{Combaluzier--Szteinsznaider:2025eoc} for their kind and helpful communication.

\section*{Acknowledgements}
This work was supported in part by World Premier International Research Center Initiative (WPI), MEXT, Japan.
The work of H.K.~was supported by JST (Japan Science and Technology Agency) SPRING, Grant No.\ JPMJSP2110.
The work of S.M.~was supported in part by JSPS (Japan Society for the Promotion of Science) KAKENHI Grant No.\ JP24K07017. 
The work of N.O.~was supported by JSPS KAKENHI Grant No.\ JP23K13111 and by the Hakubi project at Kyoto University.
The work of K.T.~was supported in part by JSPS KAKENHI Grant No.\ JP23K13101.
The work of V.Y.~was supported in part by grants for development of new faculty staff, Ratchadaphiseksomphot Fund, Chulalongkorn University and by the National Science, Research and Innovation Fund (NSRF) via the Program Management Unit for Human Resources \& Institutional Development, Research and Innovation Grant No.\ B39G680009.

\appendix

\renewcommand{\theequation}{A.\arabic{equation}}
\section*{Appendix: Homogeneous solution of odd-parity metric perturbations on a flat background}\label{App:mathematical-formula}
\addcontentsline{toc}{section}{Appendix: Homogeneous solution of odd-parity metric perturbations on a flat background}

In this appendix, we study homogeneous solutions of odd-parity metric perturbations on a flat background.
As shown in Eq.~\eqref{C0rab}, the Regge-Wheeler master variable~$\Psi_{\rm RW}$ for the odd modes is related to the Weyl tensor component~$C_{0rab}$.
In the de Donder gauge, the linearized Einstein equation, together with the Bianchi identity, implies that the Weyl tensor satisfies
\begin{align}
    \Box_4 C_{\alpha\beta\gamma\delta}=0\;,\qquad
    \partial^\mu C_{\mu\nu\alpha\beta}=0\;,
\end{align}
where $\Box_4$ is the four-dimensional d'Alembertian in the Minkowski background.
Since $C_{0rab}$ is anti-symmetric in the indices~$a$ and $b$, it can be written as $C_{0rab} = \epsilon_{ab} \chi$, where $\epsilon_{ab}$ is the completely anti-symmetric tensor on a unit two-sphere, and $\chi(t,r,\theta,\varphi)$ is a scalar function on the 2-sphere.
In terms of the transverse vector spherical harmonics~${Y_{a}^{\rm (T)}}_{\ell m}=-\frac{2}{\sqrt{\ell(\ell+1)}}{\epsilon_a}^b\nabla_b Y_{\ell m}$, which satisfy
\begin{align}
    \epsilon_{ab}Y_{\ell m}=\frac{1}{\sqrt{\ell(\ell+1)}}\nabla_{[a}{Y_{b]}^{\rm (T)}}_{\ell m}\;, \label{scalar-vector-spherical}
\end{align}
the solution for $C_{0rab}$ can be written as
\begin{align}
    C_{0rab}&=\int\frac{\de\omega}{2\pi}\sum_{\ell\ge 2}\sum_{|m|\le \ell} c_{\ell m} (\omega)e^{-i\omega t}j_\ell(\omega r)\nabla_{[a}{Y^{({\rm T})}_{b]}}_{\ell m}(\theta,\varphi)\;,\label{Weyl-source-app}
\end{align}
where $j_\ell(z)=(-1)^\ell z^\ell \left(\frac{1}{z}\frac{\de}{\de z}\right)^\ell \frac{\sin z}{z}$ is the spherical Bessel function, and $c_{\ell m}$'s are expansion coefficients depending on $\omega$.
It should be noted that we have assumed regularity of the solution in the regime~$\omega r\ll 1$.

Let us also mention the relation to the decomposition in terms of the symmetric trace-free (STF) tensors based on Ref.~\cite{Thorne:1980ru}.
The set of all STF tensors forms a $(2\ell+1)$-dimensional vector space whose basis is given by ${\mathscr{Y}}^L_{\ell m}$ (see Ref.~\cite{Thorne:1980ru} for their definition) satisfying ${\mathscr{Y}}^L_{\ell (-m)}=(-1)^m{\mathscr{Y}^*}^L_{\ell m}$.
An arbitrary STF tensor~$\mathscr{F}^L$ can thus be expanded in the basis~$\{{\mathscr{Y}}^L_{\ell m}\}$ as
\begin{equation}
\mathscr{F}^L=\sum_{|m|\le\ell}
{\mathscr{Y}^*}^L_{\ell m}\mathcal{F}_{\ell m}\;,\qquad 
\mathcal{F}_{\ell m}=\frac{4\pi\,\ell!}{(2\ell+1)!!}{\mathscr{Y}}^L_{\ell m}\mathscr{F}_L\;.
\end{equation}
Since the tensors~${\mathscr{Y}}^L_{\ell m}$ generate an irreducible representation of the rotation group of weight $\ell$, there exists a one-to-one mapping between them and the spherical harmonics~$Y_{\ell m}$.
Written explicitly, the mapping is given by
\begin{equation}
Y_{\ell m}={\mathscr{Y}^*}^L_{\ell m} n_{\langle L \rangle}\;, \qquad
{\mathscr{Y}}^L_{\ell m}= 
\frac{(2\ell+1)!!} {4\pi\,\ell!}\int\de^2\Omega\,
n^{\langle L \rangle}Y^*_{\ell m}\;,
\end{equation}
where the integration in the second equation is over a unit 2-sphere, and $n^{\langle L \rangle}\equiv n^{\langle i_1}n^{i_2}\cdots n^{i_\ell\rangle}$, with $n^i$ being the unit radial vector.
It is also useful to note that an arbitrary scalar function~$\Phi(\theta,\varphi)$ on a unit 2-sphere can be decomposed as
\begin{equation}
\Phi(\theta,\varphi)=\sum_{\ell\ge 0} \sum_{|m|\le\ell}\phi_{\ell m} Y_{\ell m}
=\sum_{\ell\ge 0} \phi^L n_{\langle L\rangle}\;,
\end{equation}
where $\phi_{\ell m}$ and $\phi_L$ are the corresponding expansion coefficients.
With these ingredients, Eq.~\eqref{Weyl-source-app} can be rewritten in the form,
\begin{align}
    C_{0rab}&=\epsilon_{ab}\int\frac{\de\omega}{2\pi}\sum_{\ell\ge 2}c^L(\omega)n_{\langle L\rangle}e^{-i\omega t}j_\ell(\omega r)\;,
\end{align}
where $c^L(\omega)\equiv \sqrt{\ell(\ell+1)}\sum_{|m|\le\ell}{\mathscr{Y}^*}^L_{\ell m}c_{\ell m}(\omega)$.

\small
\bibliographystyle{utphys}
\bibliography{bib}

\end{document}